\documentclass[twocolumn,aps,prb,groupedaddress]{revtex4-1}

\usepackage{graphicx}
\usepackage{dcolumn}
\usepackage{bm}
\usepackage{amsmath,amssymb,amsfonts,bbm}
\usepackage{color}
\usepackage{braket}

\begin{document}

\title{Enhanced anomalous Nernst effect in disordered Dirac and Weyl materials}
\author{Micha{\l} Papaj}
\author{Liang Fu}

\affiliation{Department of Physics, Massachusetts Institute of Technology, Cambridge, Massachusetts 02139, USA}

\begin{abstract}
We analyze the thermoelectric response of Dirac and Weyl semimetals using the semiclassical approach, focusing on the extrinsic contributions due to skew-scattering and side jump. Our results apply to linear response Nernst effect in ferromagnetic Dirac materials such as Fe$_3$Sn$_2$, Weyl semimetal Co$_3$Sn$_2$S$_2$ and to second order response of monolayer graphene on hBN with trigonal warping. Our analysis indicates that the extrinsic contributions can be a significant component of anomalous Nernst response and used to explain an enhanced thermoelectric response.

\end{abstract}


\maketitle

\section{Introduction}
Thermoelectricity pertains to various effects in which electric voltage appears due to the presence of a temperature gradient in a device \cite{behniaFundamentalsThermoelectricity2015, roweCRCHandbookThermoelectrics, heAdvancesThermoelectricMaterials2017, snyderComplexThermoelectricMaterials2008, zhuCompromiseSynergyHighEfficiency2017, biswasHighperformanceBulkThermoelectrics2012}. These phenomena offer a promising basis for novel devices applicable to energy conversion and cooling without the necessity for moving components, enabling their silent operation. While the progress of the field has relied mostly on the longitudinal Seebeck effects, the complementary transverse Nernst effects have been underutilized so far \cite{behniaNernstEffectMetals2016, watzmanDiracDispersionGenerates2018, fuLargeNernstPower2018}. However, such transverse device configurations have several advantages over the traditional ones. For example, they introduce a separation between the heat sources and the electrical circuitry \cite{thierschmannThreeterminalEnergyHarvester2015} or offer more versatility as they don't require combining $p$- and $n$-doped materials in one device. This resolves several outstanding issues that the Seebeck devices suffer from, such as the thermal expansion coefficient incompatibility.

However, while the presence of magnetic field can lead to interesting effects such as non-saturating thermopower \cite{skinnerLargeNonsaturatingThermopower2018} or quantized thermoelectric Hall coefficient \cite{koziiThermoelectricHallConductivity2019, fuCryogenicCoolingPower2019}, it often remains the main disadvantage of the conventional Nernst devices, being an obstacle on the path towards miniaturization and circuit integration. This issue can be resolved by instead focusing on the anomalous Nernst effect (ANE) \cite{boonaSpinCaloritronics2014, miyasatoCrossoverBehaviorAnomalous2007}, which arises in magnetic materials that break time reversal symmetry even in the absence of external field. One prominent class of materials that exhibit significant ANE are compounds with topological band structures, whose discovery has rekindled a great interest in thermoelectricity and prompted multiple experimental studies  \cite{ikhlasLargeAnomalousNernst2017, liuGiantAnomalousHall2018, sakaiGiantAnomalousNernst2018, guinZeroFieldNernstEffect2019, dingIntrinsicAnomalousNernst2019, xuLargeAnomalousNernst2019, sakaiIronbasedBinaryFerromagnets2020}. These new systems promise a significant enhancement of the observed responses due to the presence of a large Berry curvature, a fundamental ingredient of the modern band theory \cite{xiaoBerryPhaseEffects2010}. Berry curvature underlies the so-called intrinsic contribution to the anomalous response \cite{nagaosaAnomalousHallEffect2010} and has been studied as the main source of the observed effects.

However, the unavoidable presence of disorder scattering in real materials gives rise to the extrinsic contributions to the anomalous Hall and Nernst effects. The main ingredients of the extrinsic contribution are skew scattering and side jump effects. Skew scattering arises, when scattering amplitudes $w_{n,m}$ from state $m$ to state $n$ is different than the rate for the opposite process, $w_{m,n}$. One reason for breaking of this detailed balance is the presence of strong spin-orbit coupling in the system. In such a case, the scattering rate acquires an antisymmetric component, starting from the third order in the perturbative expansion in the scattering potential. The other type of extrinsic contribution is the side jump effect, which in general is due to a wave packet displacement during each of the scattering events.

It is therefore interesting to find out the contribution of these effects to the observed transverse thermoelectric response. In this work, we study the effect of skew scattering, side jump and the intrinsic contribution for two classes of the systems of interest: 2D Dirac semimetals and 3D ferromagnetic Weyl semimetals, which already have interesting longitudinal thermoelectric properties \cite{lundgrenThermoelectricPropertiesWeyl2014}. In the case of 2D Dirac semimetals, while the most celebrated example is graphene with its honeycomb lattice, since it doesn't break time reversal symmetry, the contributions from $K$ and $K'$ valley cancel each other and there is no transverse response at the linear order. Recently however, the discovery of the kagom{\'e} lattice material Fe$_3$Sn$_2$ \cite{yeMassiveDiracFermions2018} has enabled observation of the anomalous responses in its ferromagnetic state. In the case of 3D Weyl semimetals, lately there has been a spike of interest in the magnetic members of this class as they promise interesting new physics due to the interplay of magnetism and topology. One of the representatives of this category is the kagom{\'e} lattice material Co$_3$Sn$_2$S$_2$, in which large anomalous Hall and Nernst responses have been observed \cite{liuGiantAnomalousHall2018, guinZeroFieldNernstEffect2019, dingIntrinsicAnomalousNernst2019}.

Recently, there has also been an increase in interest in systems, which preserve time-reversal symmetry, but break inversion symmetry and thus can exhibit transverse response at second order \cite{sodemannQuantumNonlinearHall2015, maObservationNonlinearHall2019, duDisorderinducedNonlinearHall2019, isobeHighfrequencyRectificationChiral2020} or in ballistic conditions \cite{papajMagnusHallEffect2019}. We therefore expand our discussion of 2D Dirac semimetals by the analysis of inversion-breaking 2D Dirac semimetals, such as the monolayer graphene on hexagonal boron nitride (hBN), which can exhibit transverse response at the second order in the temperature gradient.

Motivated by all this recent materials science progress, we perform calculations using the semiclassical Boltzmann equation formalism to gain additional insight into the extrinsic contributions to the anomalous responses. We derive the formal solutions for thermoelectric contributions to the non-equilibrium electronic distribution and then apply it to the three models that approximate the materials in question. We then use the obtained formulas together with the measured material parameters such as the chemical potential or the gap size to compare the theoretical results to the transport measurements. Since the strength and number density of impurities is unknown and difficult to estimate, we use the anomalous Hall results obtained with the same formalism to make a comparison. In the case of Fe$_3$Sn$_2$ we first show that the combination of intrinsic and extrinsic contributions allows for determination of the anomalous Hall conductance within 10\% of the measured value without any fitting parameters. By applying the same set of parameters we then predict that this layered Dirac semimetal can exhibit anomalous Nernst effect of a large magnitude. In the case of Weyl semimetals, we can obtain anomalous Hall and Nernst conductivities which are close to the experimentally measured values, suggesting that the extrinsic contributions are important in considering the anomalous response in this class. Finally, by studying monolayer graphene on hBN we find that the second order ANE response in inversion-breaking breaking materials is indeed non-zero even when time reversal symmetry is preserved and is independent of temperature, opening some possibilities for low temperature applications. We also find a strong disorder strength dependence, suggesting possible significant increase in the signal strength with the sample improvements.

The remainder of the paper is organized as follows. In Section \ref{sec:formalism} we review the semiclassical Boltzmann equation formalism as applied to thermoelectric phenomena and derive the formal solutions for the Nernst response. Section \ref{sec:dirac} is devoted to ferromagnetic Dirac semimetals, with the obtained formulas applied to kagom{\'e} lattice material Fe$_3$Sn$_2$. In Section \ref{sec:weyl} we perform analogous analysis for ferromagnetic Weyl semimetals. Section \ref{sec:dirac_second} contains the expansion of the Dirac material discussion to the second order response with monolayer graphene on hBN as an example. We close with a summary and additional discussion in Section \ref{sec:summary}.

\section{\label{sec:formalism}Boltzmann formalism for Nernst effect}
\subsection{Boltzmann equation and collision integral}

In this section, we review the semiclassical Boltzmann equation formalism as applied to the thermoelectric phenomena. We consider a setup with no electric field applied, but with a constant temperature gradient throughout the whole sample $\nabla T(\mathbf{r})=\mathrm{const}$. For such a configuration we solve the Boltzmann equation:

\begin{equation}
\frac{\partial f}{\partial t} + \mathbf{\dot{r}} \cdot \frac{\partial f}{\partial \mathbf{r}} + \mathbf{\dot{k}} \cdot \frac{\partial f}{\partial \mathbf{k}} = - C[f]
\end{equation}
where $f(\mathbf{r}, \mathbf{k}, t)$ is the nonequilibrium electron distribution and $C[f]$ is the collision integral functional given by:

\begin{equation}
C[f] = \sum_{\mathbf{k}'} w_{\mathbf{k}',\mathbf{k}}f(\mathbf{r},\mathbf{k}) - w_{\mathbf{k},\mathbf{k}'}f(\mathbf{r}+\delta\mathbf{r}_{\mathbf{k}',\mathbf{k}}, \mathbf{k}')
\label{eq:collision}
\end{equation}
with $\delta\mathbf{r}_{\mathbf{k}',\mathbf{k}}$ being the real space coordinate shift that contributes to the side jump term \cite{sinitsynCoordinateShiftSemiclassical2006}. We will model the disorder as a set of randomly placed $\delta$ function potentials $V(\mathbf{r})=\sum_n V_n \delta(\mathbf{r} - \mathbf{R}_n)$, characterized by the strength with nonvanishing second and third moments $\langle V_n^2 \rangle = V_0^2, \langle V_n^3 \rangle = V_1^3$.

We supplement the Boltzmann equation with the semiclassical equations of motion of electron wave packet:
\begin{equation}
\mathbf{\dot{r}} = \frac{\partial \epsilon_\mathbf{k}}{\partial \mathbf{k}} - \mathbf{\dot{k}} \times \Omega(\mathbf{k}) +\sum_{\mathbf{k}'} w_{\mathbf{k}',\mathbf{k}} \delta\mathbf{r}_{\mathbf{k}',\mathbf{k}}, \quad \mathbf{\dot{k}} = \mathbf{F}
\end{equation}
with $\mathbf{F}$ being the force acting on the electrons. Here $\Omega(\mathbf{k})$ is the Berry curvature, which is defined as:
\begin{equation}
\mathcal{A}_a = i \langle u_\mathbf{k} | \partial_{k_a} u_\mathbf{k} \rangle, \quad \Omega_a = \epsilon_{abc} \partial_{k_b} \mathcal{A}_c.
\end{equation}
Since we are considering transport that is driven purely by the temperature gradient, $\mathbf{F}=0$. However, Berry curvature will still manifest itself at the stage of current density calculation \cite{xiaoBerryPhaseEffectAnomalous2006}.

\subsection{Scattering rates and coordinate shift}
As the focus of this work lies in the extrinsic contribution to anomalous Nernst effect, we consider elastic scattering with impurities and we use Fermi golden rule to relate the scattering rate $w_{\mathbf{k}',\mathbf{k}}$ to the $T$-matrix as follows:
\begin{equation}
w_{\mathbf{k}',\mathbf{k}} = 2 \pi |T_{\mathbf{k}',\mathbf{k}}|^2 \delta(\epsilon_{\mathbf{k}'}-\epsilon_{\mathbf{k}})
\end{equation}
The $T$-matrix can be calculated using $T_{\mathbf{k}',\mathbf{k}} = \langle \mathbf{k}'| V | \psi_\mathbf{k} \rangle$, where $| \mathbf{k} \rangle$ is an eigenstate of unperturbed Hamiltonian $H_0$ without scattering sources and $|\psi_\mathbf{k} \rangle$ is the eigenstate of the Hamiltonian with impurity potential $V$ included, which can be obtained by solving the Lippmann-Schwinger equation
\begin{equation}
|\psi_\mathbf{k} \rangle = |\mathbf{k} \rangle + (\epsilon_\mathbf{k} - H_0 + i\eta)^{-1} V |\psi_\mathbf{k} \rangle.
\end{equation}

The scattering rate $w_{\mathbf{k}',\mathbf{k}}$ can be decomposed into the symmetric and antisymmetric components:
\begin{equation}
w_{\mathbf{k}',\mathbf{k}} = w^S_{\mathbf{k}',\mathbf{k}} + w^A_{\mathbf{k}',\mathbf{k}}
\end{equation}
which by definition have the following properties:

\begin{equation}
w^S_{\mathbf{k}',\mathbf{k}} = w^S_{\mathbf{k},\mathbf{k}'}, \quad w^A_{\mathbf{k}',\mathbf{k}} = -w^A_{\mathbf{k},\mathbf{k}'}
\end{equation}

In general, the antisymmetric component is smaller than the symmetric one and its effect can be treated as a perturbation to the symmetric scattering. In the case of weak disorder, the scattering rates can be calculated as a power series in the scattering potential. The lowest nonvanishing contribution to the symmetric scattering rate appears at the order $V^2$, while the antisymmetric part begins with $V^3$ terms. We will also include antisymmetric contributions that behave as $V^4$ as they are qualitatively different from $V^3$ terms and are present even if the scattering potential distribution has a vanishing third moment.

Another significant effect that contributes to the extrinsic Nernst effect is the side jump. To account for this phenomenon, we have to consider the coordinate shift during each scattering event. This shift is independent of the impurity type and is determined only by the initial and final states in the scattering process \cite{sinitsynCoordinateShiftSemiclassical2006}:
\begin{equation}
\delta\mathbf{r}_{\mathbf{k}',\mathbf{k}} = \langle u_{\mathbf{k}'} |i\partial_{\mathbf{k}'} u_{\mathbf{k}'} \rangle - \langle u_{\mathbf{k}} |i \partial_{\mathbf{k}} u_{\mathbf{k}} \rangle - (\partial_{\mathbf{k}}+ \partial_{\mathbf{k}'}) \mathrm{arg} \langle u_{\mathbf{k}'} | u_{\mathbf{k}} \rangle
\end{equation}
where $|u_{\mathbf{k}} \rangle = \sqrt{V_\Omega} e^{-i\mathbf{k}\cdot\mathbf{r}}|\psi_\mathbf{k}\rangle$ is the cell-periodic part of the Bloch wavefunction and $\mathrm{arg}$ is the complex argument function.

\subsection{Formal solution of the Boltzmann equation}
We can now look for the formal solution of the Boltzmann equation as a perturbation to the Fermi-Dirac distribution $f_0$. As the temperature gradient is assumed to be time-independent and we are looking for the stationary state distribution, we have $\frac{\partial f}{\partial t} = 0$. The perturbative non-equilibrium distribution $f(\mathbf{r},\mathbf{k})$ can be expanded in the powers of the temperature gradient $\nabla T$ and the asymmetric scattering rate $w^A_{\mathbf{k}',\mathbf{k}}$ (as it is bound to be smaller than the symmetric component). We therefore look for a solution as the following series:
\begin{equation}
f = f_0(\mathbf{r}, \mathbf{k}) + f^\mathrm{scatt}(\mathbf{r}, \mathbf{k}) + f^\mathrm{adist}(\mathbf{r}, \mathbf{k}), 
\end{equation}
with  $f_0(\mathbf{r}, \mathbf{k}) = (1+\exp(\frac{\epsilon_\mathbf{k}-\mu}{T(\mathbf{r})}))^{-1}$ being the Fermi-Dirac distribution, $f^\mathrm{scatt}$ being due to scattering, and $f^\mathrm{adist}$ due to side jump and Berry curvature effects:
\begin{equation}
f^\mathrm{scatt}(\mathbf{r}, \mathbf{k}) = \sum_{n=1, m=0}^\infty f_n^m, \quad f^\mathrm{adist}(\mathbf{r}, \mathbf{k}) = \sum_{n=1, m=0}^\infty g_n^m
\end{equation}
In the sums above index $n$ counts the powers of $\nabla T$ and $m$ counts the powers of $w^A_{\mathbf{k}',\mathbf{k}}$. Using these definitions we can now separate the collision integral into the symmetric and antisymmetric components while also expanding the distribution in the powers of the side jump coordinate shift. This allows us to solve the Boltzmann equation iteratively order by order. In solving the equation, we define the scattering times $\tau$ as self-consistent solutions of the following equations:
\begin{equation}
\label{eq:tau}
\frac{1}{\tau_n^m} f_n^m(\mathbf{r}, \mathbf{k}) = \sum_{\mathbf{k}'} w^S_{\mathbf{k}',\mathbf{k}} (f_n^m(\mathbf{r}, \mathbf{k}) - f_n^m(\mathbf{r}, \mathbf{k}'))
\end{equation}
\begin{equation}
\label{eq:tau_p}
\frac{1}{\tau_n^{'m}} g_n^m(\mathbf{r}, \mathbf{k}) = \sum_{\mathbf{k}'} w^S_{\mathbf{k}',\mathbf{k}} (g_n^m(\mathbf{r}, \mathbf{k}) - g_n^m(\mathbf{r}, \mathbf{k}'))
\end{equation} 
The full expressions for the collision integral and the components of the perturbative solution are given in the Appendix. Equipped with these, we can now determine the thermoelectric Hall coefficient.

\subsection{Current density and thermoelectric Hall coefficient}

The formal solution to the Boltzmann equation can be used to calculate the total transport current density $\mathbf{j}$ in the system with applied temperature gradient \citep{xiaoBerryPhaseEffectAnomalous2006}:
\begin{equation}
\label{eq:current_density}
\mathbf{j} = -e \sum_\mathbf{k} \dot{\mathbf{r}} f(\mathbf{r}, \mathbf{k}) - \nabla \times T \sum_\mathbf{k} \frac{e}{\hbar} \mathbf{\Omega}(\mathbf{k}) \log \left(1+e^{-\frac{\epsilon_\mathbf{k} - \mu}{T}}\right)
\end{equation}
where the first term is responsible for the extrinsic contribution due to skew-scattering and side-jump processes, while the second term gives rise to the intrinsic contribution due to Berry curvature. At the linear order, we will be analyzing the Peltier conductivity tensor:
\begin{equation}
j_{1,a} = \alpha_{ab} (-\nabla T)_b
\end{equation}
In our discussions of both ferromagnetic Dirac and Weyl semimetals we will analyze the 3 extrinsic contributions to $\alpha_{xy}$ that arise due to impurity scattering: side-jump, third and fourth order skew scattering, and compare them with the intrinsic contribution from the Berry curvature. At lowest temperatures the intrinsic contribution can be obtained using Mott relation:
\begin{equation}
\alpha_{xy,intrinsic} = - \frac{\pi^2}{3} \frac{k_B^2 T}{|e|} \frac{d \sigma_{xy,intrinsic}}{d\epsilon}
\end{equation}
where $\sigma_{xy,intrinsic}$ is the intrinsic anomalous Hall conductivity:
\begin{equation}
\sigma_{xy,intrinsic} = -\frac{e^2}{h} \sum_\mathbf{k} f_0(\mathbf{k}) \Omega_z(\mathbf{k})
\end{equation}

The extrinsic contributions can be clearly separated from each other due to the expansion of the nonequilibrium distribution into powers of antisymmetric scattering rate and explicit split of the wavepacket velocity into the band group velocity $\mathbf{v}_0$ and the side jump velocity $\mathbf{v}_\mathrm{sj}$. We have therefore:
\begin{subequations}
\begin{align}
\alpha_{xy, \mathrm{skew}} &= e \sum_\mathbf{k} v_{0, x} f_1^1(\mathbf{r}, \mathbf{k}) /\partial_{y}T \\
\alpha_{xy, \mathrm{sj}} &= e \sum_\mathbf{k} \left(v_{0, x} g_1^0(\mathbf{r}, \mathbf{k}) + v_{sj, x} f_1^0(\mathbf{r}, \mathbf{k}) \right)/\partial_{y}T
\end{align}
\label{eq:extrinsic_contributions}
\end{subequations}
where there are two contributions to the side jump, one coming from the anomalous distribution $g_1^0(\mathbf{r}, \mathbf{k})$ and the other coming from the side jump velocity $\mathbf{v}_\mathrm{sj}$. We can further separate the skew scattering contribution by the order of the scattering potential, which we will state explicitly when discussing the particular models.
 
For the second order effects, we will investigate the $\chi_{abc}$ response tensor
\begin{equation}
j_{2,a} = \chi_{abc} (-\nabla T)_b (-\nabla T)_c.
\end{equation}
In this case, we will consider the leading order skew scattering contribution.

\section{\label{sec:dirac}Ferromagnetic Dirac semimetals}

\subsection{Model and the Nernst effect components}
To describe the thermoelectric properties of the ferromagnetic Dirac semimetals we use a simple massive Dirac fermion model:
\begin{equation}
H = \begin{pmatrix}
\Delta & v (s k_x - i k_y) \\
v (s k_x + i k_y) & -\Delta
\end{pmatrix}
\end{equation}
where $\Delta$ determines the bandgap, $v$ is the velocity of the fermion and $s=\pm 1$ is the chirality of the Dirac node, which changes under time reversal. We assume that the chemical potential is placed in the conduction band, described by the dispersion relation $\epsilon_\mathbf{k}=\sqrt{v^2 k^2+\Delta^2}$. With that assumption in mind we can proceed to calculate the symmetric and antisymmetric scattering rates. We obtain, in agreement with Ref.\citep{sinitsynAnomalousHallEffect2007}:

\begin{subequations}
\begin{align}
w^{S,2}_{\mathbf{k}',\mathbf{k}} &= \pi n_i V_0^2 \frac{\epsilon_\mathbf{k}^2 + \Delta^2 + (\epsilon_\mathbf{k}^2 - \Delta^2)\cos \left( \phi-\phi' \right)}{\epsilon_\mathbf{k}^2} \delta(\epsilon_{\mathbf{k}'}-\epsilon_\mathbf{k})\\
w^{A,3}_{\mathbf{k}',\mathbf{k}} &= - s \pi n_i V_1^3 \frac{\Delta (\epsilon_\mathbf{k}^2 - \Delta^2)\sin \left( \phi-\phi' \right)}{2 \epsilon_\mathbf{k}^2 v^2} \delta(\epsilon_{\mathbf{k}'}-\epsilon_\mathbf{k})\\
w^{A,4}_{\mathbf{k}',\mathbf{k}} &= - s \frac{3 \pi (n_i V_0^2)^2}{4} \frac{\Delta (\epsilon_\mathbf{k}^2 - \Delta^2)\sin \left( \phi-\phi' \right)}{4 |\epsilon_\mathbf{k}|^3 v^2} \delta(\epsilon_{\mathbf{k}'}-\epsilon_\mathbf{k})
\end{align}
\end{subequations}

While the symmetric scattering rate $w^{S,2}_{\mathbf{k}',\mathbf{k}}$ is independent of $s$, both antisymmetric rates contain $s$, which is expected as this means they are odd under time-reversal symmetry and their contribution to Peltier conductivity will flip under this operation. The scattering rates allow to calculate the characteristic scattering times for each order as defined in Eq. \eqref{eq:tau} and \eqref{eq:tau_p}. We obtain:

\begin{equation}
\frac{1}{\tau_1^0(\epsilon)} = \frac{1}{\tau_1^1(\epsilon)} = \frac{1}{\tau_1^{'0}(\epsilon)} =  \frac{1}{\tau_1^{'1}(\epsilon)} = n_i V_0^2 \frac{\epsilon^2_\mathbf{k} + 3\Delta^2}{4 \epsilon_\mathbf{k} v^2}.
\end{equation}
At the lowest order of scattering potential, all of the scattering times are equal and they don't have any directional dependence.

The next step is obtaining the coordinate shift for the purpose of the side jump contribution, which we find to be:

\begin{equation}
\delta \mathbf{r}_{\mathbf{k}',\mathbf{k}} = \mathbf{\Omega}(\mathbf{k}) \times \frac{\mathbf{k}'-\mathbf{k}}{|\langle u_{\mathbf{k}'} | u_\mathbf{k} \rangle|^2}, \quad \mathbf{\Omega}(\mathbf{k}) = - s \frac{\Delta v^2}{2 \epsilon_\mathbf{k}^3} \hat{\mathbf{z}}
\end{equation}

This in turn enables us to calculate the side-jump velocities, both due to symmetric and antisymmetric scattering processes as follows:

\begin{subequations}
\begin{align}
\mathbf{v}_{sj}^{S,2} &= \int \frac{d^2k'}{(2\pi)^2} w^S_{\mathbf{k}',\mathbf{k}} \delta \mathbf{r}_{\mathbf{k}',\mathbf{k}} = s \frac{n_i V_0^2 \Delta}{2 \epsilon_\mathbf{k}^2} \hat{\mathbf{z}} \times \mathbf{k}\\
\mathbf{v}_{sj}^{A,3} &= \int \frac{d^2k'}{(2\pi)^2} w^{A3}_{\mathbf{k}',\mathbf{k}} \delta \mathbf{r}_{\mathbf{k}',\mathbf{k}} = \frac{n_i V_1^3 \Delta^2 (\epsilon_\mathbf{k}^2+\Delta^2)}{4\epsilon_\mathbf{k}^2 v^3 \sqrt{\epsilon_\mathbf{k}^2-\Delta^2}}\hat{\mathbf{k}}\\
\mathbf{v}_{sj}^{A,4} &= \int \frac{d^2k'}{(2\pi)^2} w^{A4}_{\mathbf{k}',\mathbf{k}} \delta \mathbf{r}_{\mathbf{k}',\mathbf{k}} =  \frac{3 n_i^2 V_0^4 \Delta^2 (\epsilon_\mathbf{k}^2+\Delta^2)}{8\epsilon_\mathbf{k}^3 v^3 \sqrt{\epsilon_\mathbf{k}^2-\Delta^2}}\hat{\mathbf{k}}
\end{align}
\end{subequations}

Since the antisymmetric contributions to the side jump velocity are in the direction $\hat{k}$, they will not contribute to the Nernst conductivity at the lowest order, coupled to the symmetric part of the nonequilibrium distribution. On the other hand, $\mathbf{v}_{sj}^{S,2}$ is perpendicular to $\mathbf{k}$ and so will be the leading contribution to the side jump effect. With all these quantities we can now analyze each of the extrinsic contributions separately as given by the Eq.\eqref{eq:extrinsic_contributions}.

\begin{subequations}
\begin{align}
&\alpha_{xy}^\mathrm{intrinsic} = -\frac{e k_B}{h} s \frac{\pi^2 \Delta}{6 E_F^2} k_B T\\
&\alpha_{xy}^\mathrm{sj} = -\frac{e k_B}{h} s\frac{2 \pi^2 \Delta (E_F^4 - 6 E_F^2 \Delta^2 - 3 \Delta^4)}{3 E_F^2 (E_F^2 + 3 \Delta^2)^2} k_B T\\
&\alpha_{xy}^\mathrm{sk,3} = \frac{e k_B}{h} s\frac{16 \pi^2 V_1^3 \Delta^3 E_F (E_F^2 - \Delta^2)}{3 n_i V_0^4 (E_F^2 + 3 \Delta^2)^3} k_B T\\
&\alpha_{xy}^\mathrm{sk,4} = -\frac{e k_B}{h} s\frac{\pi^2 \Delta (E_F^2 - \Delta^2) (E_F^4 - 14 E_F^2 \Delta^2 - 3 \Delta^4)}{2 E_F^2 (E_F^2 + 3 \Delta^2)^3} k_B T
\end{align}
\label{eq:nernst_dirac}
\end{subequations}

We note that both of the side jump contributions are equal to each other. The most significant difference between the skew scattering contributions at third and fourth order is the dependence on the strength and concentration of the impurities: while $\alpha_{xy}^\mathrm{sk,3} \sim V_1^3/n_i V_0^4$, $\alpha_{xy}^\mathrm{sk,4}$ is independent of any of these parameters. Moreover, it doesn't require a nonvanishing third moment of the scattering potential $V_1 \neq 0$. It will therefore be also present in purely Gaussian models of disorder.

Having the explicit formulas for each of the leading order extrinsic contributions to the anomalous Nernst effect we can relate it to the measured parameters of a ferromangetic Dirac material.

\begin{figure*}
\includegraphics[width=0.92\linewidth]{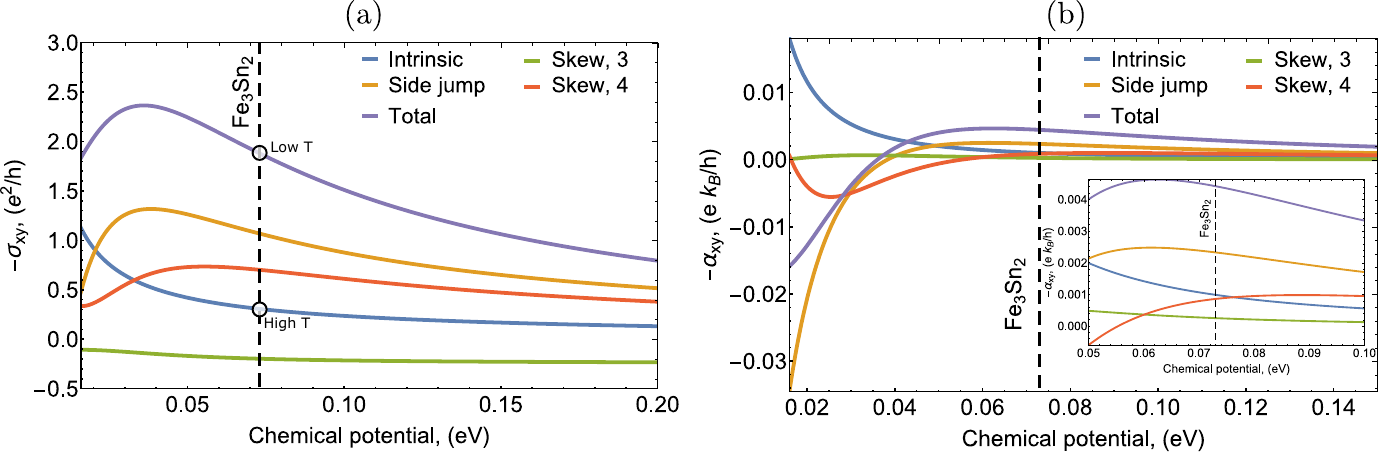}
\caption{\label{fig:Fe3Sn2} (a) The contributions to the anomalous Hall conductivity $\sigma_{xy}$ from intrinsic and extrinsic sources from Eq.\eqref{eq:hall_dirac}. The parameters are chosen such that they reproduce the measured value of $\sigma_{xy}$ for ferromagntic Dirac semimetal Fe$_3$Sn$_2$. (b) The contributions to the anomalous Nernst conductivity $\alpha_{xy}$ calculated at $T=1 \mathrm{K}$. The inset shows magnification of the vicinity of the measured chemical potential of Fe$_3$Sn$_2$.}
\end{figure*}

\subsection{Application to Fe$_3$Sn$_2$}

As an example of a ferromagnetic Dirac system we consider Fe$_3$Sn$_2$, which is a layered material with the unit cell comprised of kagome lattice bilayers interlaced with stanene layers. The band structure of the material is quasi-2D and in the plane of the bilayers it has two sets of Dirac cones that are shifted in energy, both at K and K' points of the Brillouin zone. To apply Eq.~\eqref{eq:nernst_dirac}, we therefore have to take two copies of them with energy shifted by the energy splitting of the two sets of cones $\Delta E$. Moreover, the results for a single cone should be multiplied by a factor of 2, since the Dirac cones at K and K' valleys are related by inversion symmetry and thus contribute similarly to the transport properties of the system.

For the concrete parameter values describing Fe$_3$Sn$_2$ we refer to the ARPES and quantum oscillations measurement results from  \cite{yeMassiveDiracFermions2018, yeHaasvanAlphenEffect2019a}:
\begin{gather}
E_{F1} = 73 \mathrm{meV}, \, E_{F2} = 182 \mathrm{meV}, \, v = 2.2 \cdot 10^5 \mathrm{m/s}, \notag \\  2\Delta = 32 \mathrm{meV}, \tau = 0.22 \mathrm{ps}, \, l_\mathrm{mfp} = 50 \mathrm{nm}, \, \mu_\mathrm{mob} = 800 \frac{\mathrm{cm}^2}{\mathrm{Vs}}.
\end{gather}
These parameters were determined at $T=0.6\, \mathrm{K}$. From the parameters above we can obtain the Fermi wavelength to be $\lambda_F \approx 8\, \mathrm{nm}$, so $\lambda_F < l_\mathrm{mfp}$ and the application of the semiclassical analysis to this case is justified. However, the strength and the concentration of the impurities, which is crucial to calculate the skew scattering contribution at the third order, is difficult to determine experimentally. To obtain an estimate of this quantity we can use the known semiclassical predictions for the anomalous Hall effect in the same system \cite{sinitsynAnomalousHallEffect2007}. The semiclassical $\sigma_{xy}$ has the same types of contributions as $\alpha_{xy}$ and similarly depends only on the product of $n_i V_0^4/V_1^3$:

\begin{subequations}
\begin{align}
\sigma_{xy, \mathrm{intrinsic}} &= -\frac{e^2}{h} \frac{\Delta}{2 E_F}\\
\sigma_{xy, \mathrm{sj}} &= -\frac{e^2}{h} \frac{2 \Delta (E_F^2 - \Delta^2)}{E_F (E_F^2 + 3\Delta^2)}\\
\sigma_{xy,\mathrm{skew,3}} &= -\frac{e^2}{h} \frac{V_1^3 \Delta (E_F^2 - \Delta^2)^2}{n_i V_0^4 (E_F^2 + 3 \Delta^2)^2} \\
\sigma_{xy, \mathrm{skew,4}} &= -\frac{e^2}{h} \frac{3 \Delta (E_F^2 - \Delta^2)^2}{2 E_F (E_F^2 + 3 \Delta^2)^2}
\end{align}
\label{eq:hall_dirac}
\end{subequations}

We can now relate these contributions to the experimentally determined $\sigma_{xy}$ \cite{yeMassiveDiracFermions2018}. In that experiment the measured anomalous Hall conductivity has a strong temperature dependence, with the magnitude increasing by roughly a factor of 6 when temperature is decreased. The high temperature value (converted to 2D conductivity per kagome bilayer) $\sigma_{xy,\mathrm{exp}} \approx 0.27 e^2/h$ is attributed to the intrinsic contribution. Here we can also analyze the low temperature value $\sigma_{xy,\mathrm{exp}} \approx 1.88 \, e^2/h$ and determine the parameters characterizing disorder to be $n_i V_0^4/V_1^3 = -0.265 \, \mathrm{eV}$. In fitting we assumed that this parameter is equal for all of the Dirac cones in question. This large value results in comparatively small contribution coming from skew scattering at the third order and may be due to the suppressed value of the third moment of scattering potential $V_1$ in this material. Even if we assume that the third moment of the scattering potential is absent entirely, the sum of the three remaining components is equal to $\sigma_{xy}=2.08\, e^2/h$, which is within 10\% of the low temperature experimental value, but has the benefit of having no fitting parameters. The dependence on the chemical potential in the vicinity of the measured values of the total anomalous Hall conductivity, together with all of the extrinsic contributions, is presented in Fig.\ref{fig:Fe3Sn2}(a). The two points mark the low and high temperature experimental values.

With the estimates for all the relevant parameters, we can now determine the Nernst Peltier conductivity $\alpha_{xy}$. We present all of the individual contributions as well as the total $\alpha_{xy}$ for Fe$_3$Sn$_2$ in Fig. \ref{fig:Fe3Sn2}(b). We observe that the global maximum of the total signal arises close to the band edge, where the side jump contribution dominates, overcoming the intrinsic contribution of the opposite sign. However, since Fe$_3$Sn$_2$ is a layered 3D material, changing the chemical potential via electrostatic gating is not possible and so we have to focus on the behavior close to the natural Fermi energy of the crystal. We see that we have a local maximum of total $\alpha_{xy}$ close to the measured value of $E_F$, where $\alpha_{xy} \approx 0.005 e k_B/h$ at $T= 1 \, \mathrm{K}$. It is instructive to compare this value to other materials with strong anomalous Nernst response, such as Co$_3$Sn$_2$S$_2$. Since the thickness of the Fe$_3$Sn$_2$ bilayer is about 2 nm, we can convert the theoretical prediction to a 3D $\alpha_{xy}$ value and at $T=80K$ we obtain $\alpha_{xy}^\mathrm{3D} \approx 0.67 \mathrm{A/mK}$, whereas at the same temperature Co$_3$Sn$_2$S$_2$ gives $0.83 \mathrm{A/mK}$ \cite{guinZeroFieldNernstEffect2019}. Therefore, the layered Dirac semimetal Fe$_3$Sn$_2$ is predicted to exhibit anomalous Nernst effect comparable with the largest known values among all the materials, thus making it a prospective candidate for device applications.

\section{\label{sec:weyl}Weyl semimetals}

\subsection{Model and Nernst effect components}

\begin{figure*}
\includegraphics[width=0.8\linewidth]{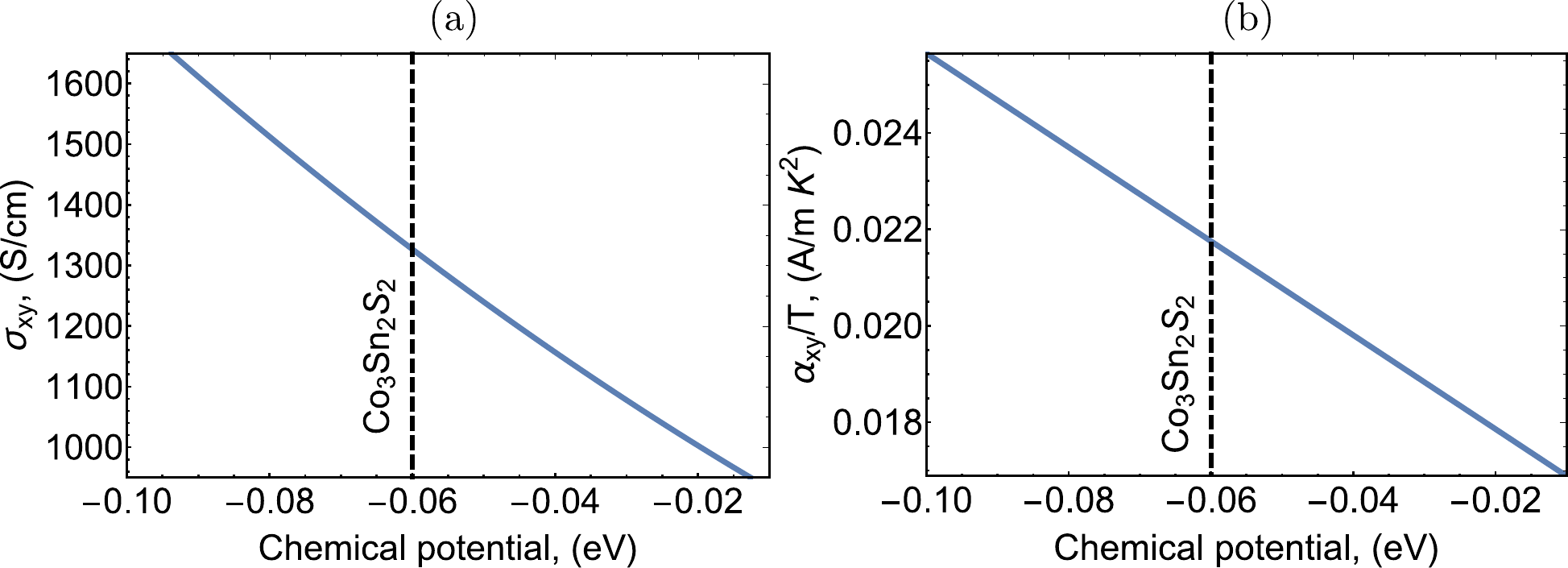}
\caption{\label{fig:Co3Sn2S2} (a) The anomalous Hall conductivity $\sigma_{xy}$ from intrinsic and extrinsic sources from Eq.\eqref{eq:hall_weyl}. The parameters are chosen such that they reproduce the measured value of $\sigma_{xy}$ for ferromagnetic Weyl semimetal Co$_3$Sn$_2$S$_2$. (b) The anomalous Nernst conductivity $\alpha_{xy}/T$ calculated for the same parameters as in (a).}
\end{figure*}

To describe Weyl semimetals we use a 3D linearly dispersing model with a tilt of the Weyl cone included.  The Hamiltonian of the model is given by:

\begin{equation}
H = s (s' v \mathbf{k} \cdot \mathbf{\sigma} + \mathbf{u}\cdot\mathbf{k})
\end{equation}
where $s=\pm 1$ is the label that distinguishes the pair of the Weyl nodes, $s'$ is the chirality of the Weyl node that changes under time-reversal, $v$ is the velocity and $\mathbf{u}=(u_x, u_y, u_z)$ are the parameters that determine the tilt, which in this form preserves inversion symmetry. The inclusion of tilt is necessary as in its absence, a single Weyl cone has an emergent time-reversal symmetry that forbids Hall or Nernst response. However, in the case of pairs of Weyl nodes that are present in real systems, untilted Weyl nodes will exhibit intrinsic anomalous Hall effect that is dependent only on the momentum space separation of the nodes \citep{burkovAnomalousHallEffect2014}. For the rest of the derivation we assume that only $u_z \neq 0$, which is the only relevant component of the tilt for $\alpha_{xy}$. We will assume that the tilt is small and obtain the results to the leading order in $u_z$. Similarly to the ferromagnetic Dirac case, we have to compute the scattering rates for the symmetric and antisymmetric processes, assuming that the third moment of disorder distribution doesn't vanish. We obtain:

\begin{subequations}
\begin{align}
w^{S,2}_{\mathbf{k}',\mathbf{k}} &= \pi  n_i V_0^2 \Bigl(1 + \sin (\theta ) \sin (\theta') \cos (\phi -\phi')  \notag \\  &+ \cos (\theta ) \cos (\theta') \Bigr) \delta(\epsilon_{\mathbf{k}'}-\epsilon_\mathbf{k})\\
w^{A,3}_{\mathbf{k}',\mathbf{k}} &= \eta s' \frac{\epsilon^2 n_i V_1^3 \sin (\theta ) \sin (\theta') \sin (\phi -\phi')}{2 v^4} u_z \delta(\epsilon_{\mathbf{k}'}-\epsilon_\mathbf{k})\\
w^{A,4}_{\mathbf{k}',\mathbf{k}} &= s' \frac{2 |\epsilon| n_i^2 V_0^4 \sin (\theta ) \sin (\theta') \sin (\phi -\phi')}{3 v^4} u_z \delta(\epsilon_{\mathbf{k}'}-\epsilon_\mathbf{k})
\end{align}
\label{eq:weyl_rates}
\end{subequations}
with $\eta =\pm1$ labeling the bands. Similarly to the Dirac case, the symmetric scattering rate is independent of Weyl node chirality, while the antisymmetric rates depend on $s'$, which means they are odd under time reversal symmetry as required. Moreover, both antisymmetric scattering rates are independent of $s$ that distinguishes the two Weyl nodes in a pair, so the total contribution of the pair doesn't vanish. We also notice that both antisymmetric rates don't have any $u_z$-independent components, so they vanish in the absence of tilt as discussed above.
With the help of the scattering rates from Eq.\eqref{eq:weyl_rates} it is now possible to determine the scattering times:
\begin{equation}
\frac{1}{\tau_1^0(\epsilon)} = \frac{1}{\tau_1^1(\epsilon)} = \frac{1}{\tau_1^{'0}(\epsilon)} =  \frac{1}{\tau_1^{'1}(\epsilon)} = \frac{\epsilon^2 n_i V_0^2}{3 \pi  v^3}
\end{equation}
Again, as in the 2D Dirac case, to the lowest order in scattering potential these scattering times are direction-independent.

Finally, to calculate the side jump contribution we obtain the coordinate shift during each scattering event:
\begin{equation}
\delta \mathbf{r}_{\mathbf{k'},\mathbf{k}} = \frac{\mathbf{\Omega}(\mathbf{k})\times(\mathbf{k}'-\mathbf{k})}{|\langle u_{\mathbf{k}'}|u_{\mathbf{k}} \rangle|^2}, \quad \mathbf{\Omega}(\mathbf{k}) = -s s' \frac{\mathbf{k}}{2 k^3}
\end{equation}
which allows us to calculate the side jump velocity:
\begin{equation}
\mathbf{v}_{sj}^{S,2} = \int \frac{d^3k'}{(2\pi)^3} w^{S,2}_{\mathbf{k}',\mathbf{k}} \delta \mathbf{r}_{\mathbf{k}',\mathbf{k}} = -s' \frac{n_i V_0^2}{2 \pi v^2} u_z \hat{\mathbf{z}} \times \mathbf{k}.
\end{equation}
In the Weyl case, the asymmetric side jump velocities will also be in the $\mathbf{k}$ direction, so they will not contribute to the leading order side jump effect. Therefore, we will focus mostly on the symmetric side jump velocity $\mathbf{v}_{sj}^{S,2}$.

With all the intermediate quantities in hand, we can turn to obtaining the extrinsic contributions to the Nernst effect in Weyl semimetals. For all the formulas below we sum over both Weyl nodes in a pair ($s=\pm 1$). By doing so we obtain:

\begin{subequations}
\begin{align}
\alpha_{xy, \mathrm{intrinsic}} &= s' \eta \frac{e k_B}{\hbar^2} \frac{u_z}{18 v^2} k_B T\\
\alpha_{xy, \mathrm{sj}} &= -s' \eta \frac{e k_B}{\hbar^2} \frac{u_z}{3 v^2} k_B T\\
\alpha_{xy,\mathrm{skew,3}} &= -s' \eta \frac{e k_B}{\hbar^2} \frac{u_z }{6 v^2} \frac{E_F V_1^3}{n_i V_0^4}k_B T\\
\alpha_{xy,\mathrm{skew,4}} &= -s' \eta \frac{e k_B}{\hbar^2} \frac{u_z}{9 v^2} k_B T
\end{align}
\label{eq:nernst_weyl}
\end{subequations}
The intrinsic contribution that we obtained from the semiclassical analysis is consistent with the results from the current-current correlation function from Ref.\cite{ferreirosAnomalousNernstThermal2017}. We notice that all of the contributions apart from the skew scattering at the third order are independent of the chemical potential and are mostly given by the magnitude of the tilting of the cone $u_z/v$. However, the skew scattering at the third order is also determined by the ratio of Fermi energy $E_F$ and the combined potential strength and concentration of the impurities $n_i V_0^4/V_1^3$. Since most of the Weyl semimetals contain several pairs of Weyl nodes, in general it can lead to a complicated overlap of all these contributions if the Weyl nodes are at several different energies.

\subsection{Estimated magnitude of anomalous response in Weyl semimetal}
With all the components of the Nernst effect for 3D Weyl semimetals calculated, we can now estimate the magnitude of the effect in a real material from this class. Similarly to the ferromagnetic Dirac semimetal, we can use the independent measurements of the anomalous Hall conductivity to better analyze our semiclassical calculations and estimate the impurity strength and concentration as the Nernst effect. The contributions to the total $\sigma_{xy}$ in this case are:
\begin{subequations}
\begin{align}
\sigma_{xy, \mathrm{intrinsic}} &=s' \frac{e^2}{h} \frac{Q}{2\pi} - s' \eta \frac{e^2}{\hbar^2} \frac{u_z E_F}{6 \pi^2 v^2} \\
\sigma_{xy, \mathrm{sj}} &= s' \eta \frac{e^2}{\hbar^2} \frac{u_z E_F}{\pi^2 v^2} \\
\sigma_{xy,\mathrm{skew,3}} &= s' \eta \frac{e^2}{\hbar^2} \frac{u_z }{4 \pi^2 v^2} \frac{E_F^2 V_1^3}{n_i V_0^4}\\
\sigma_{xy,\mathrm{skew,4}} &= s' \eta \frac{e^2}{\hbar^2} \frac{u_z E_F}{3 \pi^2 v^2}
\end{align}
\label{eq:hall_weyl}
\end{subequations}
where $Q$ is the distance between the pair of the Weyl nodes.

We can now use these formulas to estimate the response in a Weyl semimetal. One example of such a material is Co$_3$Sn$_2$S$_2$, which breaks time reversal symmetry. It has recently attracted a significant attention as it exhibits large anomalous Hall and Nernst effects \cite{dingIntrinsicAnomalousNernst2019, liuGiantAnomalousHall2018, moraliFermiarcDiversitySurface2019, guinZeroFieldNernstEffect2019}. ARPES measurements and first-principles calculations reveal that it possesses three pairs of Weyl nodes in its Brillouin zone, which are relatively close to the Fermi energy. Based on the first principles calculations \cite{liuGiantAnomalousHall2018}, we can estimate the relevant parameters to be $E_F = -60\,\mathrm{meV}$, $Q\approx 4.7 \mathrm{nm}^{-1}$ and $v \approx 10^5 \mathrm{m/s}$. Plugging in these values for 3 Weyl pairs of Co$_3$Sn$_2$S$_2$ and taking the tilt parameter to be $u_z/v=0.5$ we obtain $\sigma_{xy} \approx 1330 \,\mathrm{S/cm}$ when $n_i V_0^4/V_1^3\approx -0.07\, \mathrm{eV}$. This estimate $\sigma_{xy}$ lies within the range of measured values of $1100 - 1350\, \mathrm{S/cm}$ \cite{dingIntrinsicAnomalousNernst2019}. We present a sum of all these contributions with the parameters listed above in Fig.~\ref{fig:Co3Sn2S2}(a). We now apply the obtained scattering characterization to determine the anomalous Nernst effect. With the same set of parameters as for the anomalous Hall effect we plot the $\alpha_{xy}/T$ contributions using Eq.\ref{eq:nernst_weyl} in Fig.~\ref{fig:Co3Sn2S2}(b). For the previously taken chemical potential value corresponding to that of Co$_3$Sn$_2$S$_2$, we obtain $\alpha_{xy}/T\approx 0.022\, \mathrm{A/mK^2}$. In the experiment, the observed behavior also varies between the samples and the measured linear coefficient lies within $0.04 - 0.1 \mathrm{A/(m K^2)}$ range \cite{dingIntrinsicAnomalousNernst2019}. While there is a mismatch between the calculated and measured values, the extrinsic contributions are still larger than the purely intrinsic result. Moreover, the most important extrinsic contributions are independent of the impurity strength and concentration at the lowest order, consistent with the observation that significant changes in mobility do not impact Nernst effect strongly \cite{dingIntrinsicAnomalousNernst2019}. This suggests that the extrinsic contribution may play an important role in fully understanding the observed values of $\alpha_{xy}$ in Co$_3$Sn$_2$S$_2$, but it still requires more detailed modeling that takes into account realistic band structure and scattering model. 

\section{\label{sec:dirac_second}Inversion-breaking 2D Dirac semimetal}

\subsection{Model and the second order response}

Both monolayer graphene on hBN and bilayer graphene with perpendicular electric field can be described by the same low energy model with the Hamiltonian:
\begin{equation}
H_s = \begin{pmatrix}
\Delta & s v k_{-s} - \lambda k_s^2 \\
s v k_{s} - \lambda k_{-s}^2 & -\Delta
\end{pmatrix}
\end{equation}
Here $k_\pm = k_x \pm i k_y$,  $\Delta$ opens up the gap in the spectrum, $v$ characterizes the velocity of the Dirac cone and $\lambda$ determines the strength of the trigonal warping. While the Hamiltonian in this case differs from the ferromagnetic Dirac semimetal case by the addition of the trigonal warping only, we have to take into account that due to the time reversal symmetry there will always be a pair of Dirac cones with opposite sign of $s$ in the Brillouin zone, which we have to sum over. As we will see, this will lead to the vanishing of the first order response, while the second order contribution will still be present. We again set the chemical potential in the conduction band, which is described by the dispersion:
\begin{equation}
\epsilon_\mathbf{k} = \sqrt{v^2 k^2 + \Delta^2 + \lambda^2 k^4 - 2s v \lambda k^3 \cos(3\phi)}.
\end{equation}

\subsection{Application to monolayer graphene on hBN}

In the case of the monolayer graphene, $v$ term in the Hamiltonian dominates over the $\lambda$ term, so we can do the calculations perturbatively to the first order in $\lambda$. We begin, similarly to the previous two cases by determining the symmetric and antisymmetric scattering rates:

\begin{subequations}
\begin{align}
&w^{S,2}_{\mathbf{k}',\mathbf{k}} = \pi n_i V_0^2 \delta(\epsilon_{\mathbf{k}'}-\epsilon_\mathbf{k})\Bigl(\frac{\epsilon_\mathbf{k}^2 + \Delta^2 + (\epsilon_\mathbf{k}^2 - \Delta^2)\cos \left( \phi-\phi' \right)}{\epsilon_\mathbf{k}^2}  \notag \\  &+ \frac{s \lambda(\epsilon_\mathbf{k}^2-\Delta^2)^{3/2}(\cos(\phi-4\phi')-\cos(\phi+2\phi')+(\phi \leftrightarrow \phi'))}{2 v^2\epsilon_\mathbf{k}^2}\Bigr) \\
&w^{A,3}_{\mathbf{k}',\mathbf{k}} = - \pi n_i V_1^3 \Delta \delta(\epsilon_{\mathbf{k}'}-\epsilon_\mathbf{k}) \Bigl( \frac{s (\epsilon_\mathbf{k}^2 - \Delta^2)\sin \left( \phi-\phi' \right)}{2 \epsilon_\mathbf{k}^2 v^2} \notag \\&+ \frac{ \lambda(\epsilon_\mathbf{k}^2-\Delta^2)^{3/2}(\sin(\phi-4\phi')-\sin(\phi+2\phi')-(\phi \leftrightarrow \phi'))}{4 \epsilon_\mathbf{k}^2 v^4} \Bigr) \\
&w^{A,4}_{\mathbf{k}',\mathbf{k}} = - \frac{3 \pi (n_i V_0^2)^2}{4} \Delta \delta(\epsilon_{\mathbf{k}'}-\epsilon_\mathbf{k}) \Bigl( \frac{s(\epsilon_\mathbf{k}^2 - \Delta^2)\sin \left( \phi-\phi' \right)}{|\epsilon_\mathbf{k}|^3 v^2} \notag \\&+ \frac{\lambda(\epsilon_\mathbf{k}^2-\Delta^2)^{3/2}(\sin(\phi-4\phi')-\sin(\phi+2\phi')-(\phi \leftrightarrow \phi'))}{2|\epsilon_\mathbf{k}|^3 v^4} \Bigr) 
\end{align}
\end{subequations}
where $(\phi \leftrightarrow \phi')$ means additional two terms as the ones in the same bracket with $\phi$ and $\phi'$ interchanged, which signifies the symmetric or antisymmetric character of the scattering rate. We note that in addition to the terms already present in the calculations shown in Section \ref{sec:dirac}, we obtain additional contributions due to trigonal warping. Crucially, we observe their opposite dependence on the valley parameter $s$: while in symmetric rate the original part is $s$-indepedent and $\lambda$ term switches sign between valleys, in the antisymmetric rates the $\lambda$ term is the same for both valleys and the initial term switches sign. This is the basis of vanishing linear order response and non-vanishing second order response.

Equipped with the scattering rates, we can calculate the additional scattering times for the second order processes:

\begin{equation}
\frac{1}{\tau_2^0(\epsilon)} = n_i V_0^2 \frac{\epsilon^2 + \Delta^2}{2 \epsilon v^2}, \quad \frac{1}{\tau_2^1(\epsilon)} = n_i V_0^2 \frac{\epsilon^2 + 3\Delta^2}{4 \epsilon v^2}
\end{equation}
We only keep the lowest non-vanishing order in this calculation to keep the scattering times direction-independent.

We also calculate the coordinate shift during the scattering events and integrate it with the symmetric scattering rate to obtain the side jump velocity:
\begin{equation}
\mathbf{v}_{sj}^{S,2} = \frac{s n_i V_0^2}{8 v \epsilon_\mathbf{k}^2}  \left(v k(\Delta-\epsilon_\mathbf{k}) + \sqrt{\epsilon_\mathbf{k}^2-\Delta^2}(\epsilon_\mathbf{k}+3\Delta)\right) \hat{\mathbf{z}} \times \hat{\mathbf{k}}
\end{equation}

With all these quantities at hand, we move on to calculate the transport coefficients. First, by referring to the Eq.\eqref{eq:nernst_dirac} we notice that all of the linear response contributions are multiplied by the valley index $s$. This means that the linear Nernst response vanishes when summed over both graphene valleys as it should in time-reversal invariant system. Therefore, we have to turn to the second order response. To begin with, we note that the intrinsic response at the second order depends on the Berry curvature dipole \cite{sodemannQuantumNonlinearHall2015}. However, in the present model there is no dipole moment of the Berry curvature distribution and so the intrinsic contribution vanishes. Similarly, the side jump contributions from the side jump velocity and the anomalous distribution cancel each other out. This means that the only non-vanishing components are those given by skew scattering. These are given by:

\begin{subequations}
\label{eq:chi_monolayer}
\begin{align}
\chi^\mathrm{sk,a}_{yxx} &= \frac{2 \pi e v \Delta (E_F^2-\Delta^2)^2}{(E_F^2 + \Delta^2)^2 (E_F^2+3\Delta^2)^4} \lambda \chi_a \\
\chi_3 &= \frac{8 V_1^3 E_F \Delta^2 (17 E_F^4+ 46 E_F^2 \Delta^2 + 33 \Delta^4)}{3 n_i^2 V_0^6} \\
\chi_4 &=  \frac{-3 E_F^8+54E_F^6\Delta^2+172E_F^4\Delta^4 + 146E_F^2\Delta^6+15\Delta^8}{n_i V_0^2 E_F^2}
\end{align}
\end{subequations}

From the calculated value of $\chi_{yxx}$ we see that the second order response coefficient is independent of temperature, in contrast to the linear response $\alpha_{xy}$, which vanishes linearly as $T\rightarrow 0$. This may be advantageous for applications in the low temperature regime. We also note a very strong dependence on the impurity strength and concentration, so the observed effect can be significantly changed by improving the sample quality.

\begin{figure}
\includegraphics[width=0.99\linewidth]{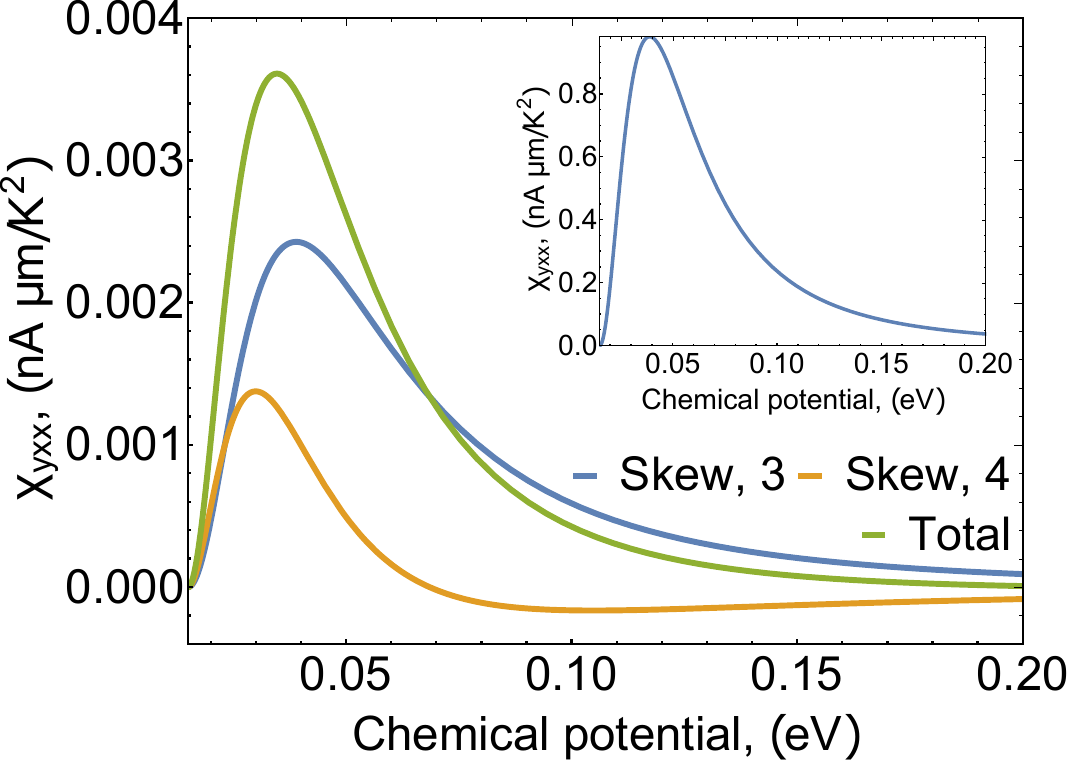}
\caption{\label{fig:monolayer_2order} The second order response tensor coefficient $\chi_{yxx}$ calculated from Eq.\eqref{eq:chi_monolayer} for the monolayer graphene on hBN.}
\end{figure}

In the case of the second order response in monolayer graphene on hBN we are unaware of any experimental measurements that would allow simultaneous determination of anomalous Hall and Nernst contributions and finding the various moments of impurity scattering potential. We therefore provide only some estimates using similar parameters and focusing on the primary (as opposed to the superlattice) Dirac points as in Ref.\cite{isobeHighfrequencyRectificationChiral2020} and so we have:
\begin{gather}
a = 0.142\mathrm{nm}, \quad v = 10^6 \mathrm{m/s}, \quad \lambda = v a / 4, \notag \\ n_i = 10^{13} \mathrm{m^{-2}}, V_0 = 2.77 \cdot 10^{-17} \mathrm{eV m^2}, \Delta = 0.015 \mathrm{eV}.
\end{gather}
We present both skew scattering components and their total in Fig. \ref{fig:monolayer_2order} when $V_1 = V_0/4$ so that both third and fourth order contribution have a comparable magnitude. We also show the alternative scenario with $V_1 = V_0 = 1.5 \cdot 10^{-17} \mathrm{eV m^2}$ in the inset, when the third order contribution dominates. We can compare this value to the ferromagnetic Dirac case by assuming a temperature gradient of $\nabla T = 1 \mathrm{K/\mu m}$, attainable in experimental studies of Nernst effect \cite{xuLargeAnomalousNernst2019}. In such a scenario, the effective $\alpha_{xy}^\mathrm{eff} = \chi_{yxx} \nabla T$ ranges from 0.001 $e k_B/h$ for strong disorder to about 0.3 $e k_B/h$ for the cleaner sample parameters, compared to 0.004 $e k_B/h$ for Fe$_3$Sn$_2$ at T=1K. This means that at low temperatures and for clean samples, the second order response has a potential of having a significant magnitude.

\section{\label{sec:summary}Summary and outlook}
In summary, in this work we investigated the extrinsic contributions to the anomalous Nernst effect in 2D Dirac and 3D Weyl semimetals. By performing semiclasical analysis in the Boltzmann equation formalism we have obtained the linear response components of $\alpha_{xy}$ that arise due to the side jump and skew scattering, together with the intrinsic contribution due to the Berry curvature.  We applied the obtained formulas to analyze the response of two materials that have attracted significant attention in the context of their anomalous responses, ferromagnetic Dirac semimetal Fe$_3$Sn$_2$ and Weyl semimetal Co$_3$Sn$_2$S$_2$. In the case of Fe$_3$Sn$_2$, we found that the semiclassical expressions precisely describe the observed anomalous Hall response based solely on the experimentally obtained band structure parameters. We then used this knowledge to predict the Nernst response to be comparable to the leading material platforms for thermoelectricity. In the case of Weyl semimetals, we have obtained an estimate of the anomalous Hall and Nernst response which are comparable to the values observed so far in Co$_3$Sn$_2$S$_2$, suggesting that extrinsic effects may be a significant contribution to the measured values. Finally, we have also investigated the second order thermoelectric response in materials that break inversion symmetry, but still preserve time-reversal symmetry, studying monolayer graphene on hBN as an example. In contrast to the linear response the second order contributions do not vanish and are independent of temperature in the low temperature limit, indicating potential applications in this regime. The magnitude of the response is also greatly dependent on the impurity concentration and strength, promising a significant enhancement with the sample quality improvements. 

Our analysis provides an indication that the extrinsic effects are an important contribution to the anomalous Nernst effect and their impact should be taken into account when considering new materials for potential applications. However, this study is only a first step on the path towards accurate quantitative description of anomalous Nernst effect in particular materials. One evident improvement to our discussion would necessitate including proper, first principles band structures in calculation of both the intrinsic contribution due to the Berry curvature and the scattering rates. Our study doesn't also take into account the impact of phonons and the effects of interactions. Nevertheless, we believe that this work will underlie some future studies that will take these effects into consideration and will make possible material design for optimal thermoelectric materials.

\begin{acknowledgments}
This work is supported by DOE Office of Basic Energy Sciences under Award DE-SC0018945.
\end{acknowledgments}

\appendix
\begin{widetext}
\section{The expressions for scattering rates}

Throughout the text for all the models under consideration we calculate the scattering rates using the Born approximation, retaining the lowest order non-vanishing contributions. The general expressions for the symmetric and antisymmetric scattering rates are:
\begin{align}
&w^{S,2}_{\mathbf{k}',\mathbf{k}} = 2\pi \langle |V_{\mathbf{k}',\mathbf{k}}|^2 \rangle\delta(\epsilon_{\mathbf{k}'}-\epsilon_{\mathbf{k}}) \\
&w^{A,3}_{\mathbf{k}',\mathbf{k}} = -(2\pi)^2 \sum_\mathbf{q} \mathrm{Im} \langle V_{\mathbf{k}',\mathbf{q} }V_{\mathbf{q},\mathbf{k}} V_{\mathbf{k},\mathbf{k}'} \rangle\delta(\epsilon_{\mathbf{k}'}-\epsilon_{\mathbf{k}})  \delta(\epsilon_{\mathbf{k}'}-\epsilon_{\mathbf{q}})\\
&w^{A,4}_{\mathbf{k}',\mathbf{k}} = 2\pi \sum_\mathbf{q} \Bigl(   \frac{1}{\epsilon_\mathbf{q}^+ - \epsilon_\mathbf{q}^-} \mathrm{Im} \langle V_{\mathbf{k}',\mathbf{q}}^{++} V_{\mathbf{q},\mathbf{k}'}^{-+} \rangle \langle V_{\mathbf{q},\mathbf{k}}^{++} V_{\mathbf{k},\mathbf{q}}^{+-} \rangle  \notag \\ &-\frac{1}{\epsilon_\mathbf{k}^+ - \epsilon_\mathbf{k}^-} \mathrm{Im} \langle V_{\mathbf{k}',\mathbf{k}}^{++} V_{\mathbf{k},\mathbf{k}'}^{-+} \rangle \langle V_{\mathbf{q},\mathbf{k}}^{+-} V_{\mathbf{k},\mathbf{q}}^{++} \rangle \notag \\ &- \frac{1}{\epsilon_{\mathbf{k}'}^+ - \epsilon_{\mathbf{k}'}^-} \mathrm{Im} \langle V_{\mathbf{k}',\mathbf{k}}^{++} V_{\mathbf{k},\mathbf{k}'}^{+-} \rangle \langle V_{\mathbf{q},\mathbf{k}'}^{++} V_{\mathbf{k}',\mathbf{q}}^{-+} \rangle \Bigr)   \delta(\epsilon_{\mathbf{k}'}-\epsilon_{\mathbf{k}})  \delta(\epsilon_{\mathbf{k}'}-\epsilon_{\mathbf{q}})
\end{align}

\section{The collision integral}

We can obtain the series expansion of Eq.\eqref{eq:collision} in the perturbative components of the non-equilibrium electronic distribution, grouping terms by the powers of dependence on the temperature gradient and the antisymmetric scattering rates:

\begin{align}
C[f] &= \int \frac{d^2k'}{(2\pi)^2} w^S_{\mathbf{k}',\mathbf{k}} \biggl[-\frac{\partial f_0}{\partial \mathbf{r}} \cdot \delta\mathbf{r}_{\mathbf{k}'\mathbf{k}} - \frac{1}{2}\frac{\partial f_0}{\partial r_a}\frac{\partial f_0}{\partial r_b}(\delta\mathbf{r}_{\mathbf{k}'\mathbf{k}})_a (\delta\mathbf{r}_{\mathbf{k}'\mathbf{k}})_b + \notag \\  &+ \sum_{nm} \left(f_n^m(\mathbf{r}, \mathbf{k}) - f_n^m(\mathbf{r}, \mathbf{k}') - \frac{\partial f_n^m}{\partial \mathbf{r}} \cdot \delta\mathbf{r}_{\mathbf{k}'\mathbf{k}} + g_n^m(\mathbf{r}, \mathbf{k}) - g_n^m(\mathbf{r}, \mathbf{k}') -\frac{\partial g_n^m}{\partial \mathbf{r}} \cdot \delta\mathbf{r}_{\mathbf{k}'\mathbf{k}} \right) \biggr] + \notag \\ &+ w^A_{\mathbf{k}',\mathbf{k}} \biggl[ f_0(\mathbf{r}, \mathbf{k}') + \frac{\partial f_0}{\partial \mathbf{r}} \cdot \delta\mathbf{r}_{\mathbf{k}'\mathbf{k}} + \sum_{nm} \left(f_n^m(\mathbf{r}, \mathbf{k}') + g_n^m(\mathbf{r}, \mathbf{k}') + \left(\frac{\partial f_n^m}{\partial \mathbf{r}} + \frac{\partial g_n^m}{\partial \mathbf{r}}\right) \cdot \delta\mathbf{r}_{\mathbf{k}'\mathbf{k}} \right) \biggr]
\end{align}

\section{Components of the formal solution of Boltzmann equation}

By following the procedure outlined in Section \ref{sec:formalism} we can formally solve the Boltzmann equation in powers of the temperature gradient (lower index) and the antisymmetric scattering rates (upper index). We obtain the following components that are used to compute the transport coefficients for all the models:

\begin{align}
f_1^0(\mathbf{r}, \mathbf{k}) &= \tau_1^0(\epsilon_\mathbf{k} - \mu) \frac{\partial f_0}{\partial \epsilon} \mathbf{\dot{r}}\cdot \frac{\nabla T}{T} \\ 
f_1^1(\mathbf{r}, \mathbf{k}) &= -\tau_1^1 \tau_1^0 \int_{\mathbf{k}'} w^A_{\mathbf{k}',\mathbf{k}} (\epsilon_{\mathbf{k}'} - \mu) \frac{\partial f_0}{\partial \epsilon} \mathbf{\dot{r}}\cdot \frac{\nabla T}{T} \\
g_1^0(\mathbf{r}, \mathbf{k}) &= - \tau_1^{'0} \int_{\mathbf{k}'} w^S_{\mathbf{k}',\mathbf{k}} (\epsilon_{\mathbf{k}'} - \mu) \frac{\partial f_0}{\partial \epsilon} \delta\mathbf{r_{\mathbf{k}',\mathbf{k}}}\cdot \frac{\nabla T}{T} \\
g_1^1(\mathbf{r}, \mathbf{k}) &= \tau_1^{'1} \int_{\mathbf{k}'} w^A_{\mathbf{k}',\mathbf{k}} \left[ (\epsilon_{\mathbf{k}'} - \mu) \frac{\partial f_0}{\partial \epsilon} \delta\mathbf{r_{\mathbf{k}',\mathbf{k}}}\cdot \frac{\nabla T}{T} + \tau_1^0  \int_{\mathbf{k}''} w^S_{\mathbf{k}'',\mathbf{k}'} (\epsilon_{\mathbf{k}''} - \mu) \frac{\partial f_0}{\partial \epsilon} \delta\mathbf{r_{\mathbf{k}'',\mathbf{k}'}}\cdot \frac{\nabla T}{T} \right]\\
f_2^0(\mathbf{r}, \mathbf{k}) &= \tau_2^0 \tau_1^0(\epsilon_\mathbf{k} - \mu) \left((\epsilon_\mathbf{k} - \mu) \frac{\partial^2 f_0}{\partial \epsilon^2} \left(\mathbf{\dot{r}}\cdot \frac{\nabla T}{T}\right)^2 + \frac{\dot{r}_a \dot{r}_b}{T^2} \frac{\partial T}{\partial r_a} \frac{\partial T}{\partial r_b} \frac{\partial f_0}{\partial \epsilon}\right)  \\
f_2^1(\mathbf{r}, \mathbf{k}) &= -\tau_2^1 \tau_1^0 (\tau_2^0 + \tau_1^1) \int_{\mathbf{k}'} w^A_{\mathbf{k}',\mathbf{k}} \left[ (\epsilon_{\mathbf{k}'} - \mu) \left((\epsilon_{\mathbf{k}'} - \mu) \frac{\partial^2 f_0}{\partial \epsilon^2} \left(\mathbf{\dot{r}}\cdot \frac{\nabla T}{T}\right)^2 + \frac{\dot{r}_a \dot{r}_b}{T^2} \frac{\partial T}{\partial r_a} \frac{\partial T}{\partial r_b} \frac{\partial f_0}{\partial \epsilon}\right) \right]
\end{align}

where the integration over momentum depends on the dimensionality of the model $\int_{\mathbf{k}} \rightarrow \int d^dk/(2\pi)^d$.

In evaluating the integrals in the low temperature limit for the first and second order contributions, we use the Sommerfeld expansion:
\begin{align}
&\int_{-\infty}^\infty dE H(E) \left(-\frac{\partial f_0}{\partial E}\right) = H(E_F) + \frac{\pi^2 T^2}{6} \left. \frac{d^2 H}{dE^2}\right|_{E=E_F} + ... \\
&\int_{-\infty}^\infty dE H(E) \frac{\partial^2 f_0}{\partial E^2} = \left. \frac{dH}{dE} \right|_{E=E_F} + \frac{\pi^2 T^2}{6} \left. \frac{d^3 H}{dE^3}\right|_{E=E_F} + ...
\end{align}
\end{widetext}

\bibliography{thermoelectric_Nernst}

\begin{thebibliography}{38}%
\makeatletter
\providecommand \@ifxundefined [1]{%
 \@ifx{#1\undefined}
}%
\providecommand \@ifnum [1]{%
 \ifnum #1\expandafter \@firstoftwo
 \else \expandafter \@secondoftwo
 \fi
}%
\providecommand \@ifx [1]{%
 \ifx #1\expandafter \@firstoftwo
 \else \expandafter \@secondoftwo
 \fi
}%
\providecommand \natexlab [1]{#1}%
\providecommand \enquote  [1]{``#1''}%
\providecommand \bibnamefont  [1]{#1}%
\providecommand \bibfnamefont [1]{#1}%
\providecommand \citenamefont [1]{#1}%
\providecommand \href@noop [0]{\@secondoftwo}%
\providecommand \href [0]{\begingroup \@sanitize@url \@href}%
\providecommand \@href[1]{\@@startlink{#1}\@@href}%
\providecommand \@@href[1]{\endgroup#1\@@endlink}%
\providecommand \@sanitize@url [0]{\catcode `\\12\catcode `\$12\catcode
  `\&12\catcode `\#12\catcode `\^12\catcode `\_12\catcode `\%12\relax}%
\providecommand \@@startlink[1]{}%
\providecommand \@@endlink[0]{}%
\providecommand \url  [0]{\begingroup\@sanitize@url \@url }%
\providecommand \@url [1]{\endgroup\@href {#1}{\urlprefix }}%
\providecommand \urlprefix  [0]{URL }%
\providecommand \Eprint [0]{\href }%
\providecommand \doibase [0]{http://dx.doi.org/}%
\providecommand \selectlanguage [0]{\@gobble}%
\providecommand \bibinfo  [0]{\@secondoftwo}%
\providecommand \bibfield  [0]{\@secondoftwo}%
\providecommand \translation [1]{[#1]}%
\providecommand \BibitemOpen [0]{}%
\providecommand \bibitemStop [0]{}%
\providecommand \bibitemNoStop [0]{.\EOS\space}%
\providecommand \EOS [0]{\spacefactor3000\relax}%
\providecommand \BibitemShut  [1]{\csname bibitem#1\endcsname}%
\let\auto@bib@innerbib\@empty
\bibitem [{\citenamefont
  {Behnia}(2015)}]{behniaFundamentalsThermoelectricity2015}%
  \BibitemOpen
  \bibfield  {author} {\bibinfo {author} {\bibfnamefont {K.}~\bibnamefont
  {Behnia}},\ }\href@noop {} {\emph {\bibinfo {title} {Fundamentals of
  {{Thermoelectricity}}}}}\ (\bibinfo  {publisher} {{Oxford University
  Press}},\ \bibinfo {address} {{Oxford, New York}},\ \bibinfo {year}
  {2015})\BibitemShut {NoStop}%
\bibitem [{\citenamefont {Rowe}()}]{roweCRCHandbookThermoelectrics}%
  \BibitemOpen
  \bibfield  {author} {\bibinfo {author} {\bibfnamefont {D.}~\bibnamefont
  {Rowe}},\ }\href@noop {} {\enquote {\bibinfo {title} {{{CRC Handbook}} of
  {{Thermoelectrics}}},}\ }\bibinfo {howpublished}
  {https://www.crcpress.com/CRC-Handbook-of-Thermoelectrics/Rowe/p/book/9780849301469}\BibitemShut
  {NoStop}%
\bibitem [{\citenamefont {He}\ and\ \citenamefont
  {Tritt}(2017)}]{heAdvancesThermoelectricMaterials2017}%
  \BibitemOpen
  \bibfield  {author} {\bibinfo {author} {\bibfnamefont {J.}~\bibnamefont
  {He}}\ and\ \bibinfo {author} {\bibfnamefont {T.~M.}\ \bibnamefont {Tritt}},\
  }\href {\doibase 10.1126/science.aak9997} {\bibfield  {journal} {\bibinfo
  {journal} {Science}\ }\textbf {\bibinfo {volume} {357}} (\bibinfo {year}
  {2017}),\ 10.1126/science.aak9997}\BibitemShut {NoStop}%
\bibitem [{\citenamefont {Snyder}\ and\ \citenamefont
  {Toberer}(2008)}]{snyderComplexThermoelectricMaterials2008}%
  \BibitemOpen
  \bibfield  {author} {\bibinfo {author} {\bibfnamefont {G.~J.}\ \bibnamefont
  {Snyder}}\ and\ \bibinfo {author} {\bibfnamefont {E.~S.}\ \bibnamefont
  {Toberer}},\ }\href {\doibase 10.1038/nmat2090} {\bibfield  {journal}
  {\bibinfo  {journal} {Nature Mater}\ }\textbf {\bibinfo {volume} {7}},\
  \bibinfo {pages} {105} (\bibinfo {year} {2008})}\BibitemShut {NoStop}%
\bibitem [{\citenamefont {Zhu}\ \emph {et~al.}(2017)\citenamefont {Zhu},
  \citenamefont {Liu}, \citenamefont {Fu}, \citenamefont {Heremans},
  \citenamefont {Snyder},\ and\ \citenamefont
  {Zhao}}]{zhuCompromiseSynergyHighEfficiency2017}%
  \BibitemOpen
  \bibfield  {author} {\bibinfo {author} {\bibfnamefont {T.}~\bibnamefont
  {Zhu}}, \bibinfo {author} {\bibfnamefont {Y.}~\bibnamefont {Liu}}, \bibinfo
  {author} {\bibfnamefont {C.}~\bibnamefont {Fu}}, \bibinfo {author}
  {\bibfnamefont {J.~P.}\ \bibnamefont {Heremans}}, \bibinfo {author}
  {\bibfnamefont {J.~G.}\ \bibnamefont {Snyder}}, \ and\ \bibinfo {author}
  {\bibfnamefont {X.}~\bibnamefont {Zhao}},\ }\href {\doibase
  10.1002/adma.201605884} {\bibfield  {journal} {\bibinfo  {journal} {Advanced
  Materials}\ }\textbf {\bibinfo {volume} {29}},\ \bibinfo {pages} {1605884}
  (\bibinfo {year} {2017})}\BibitemShut {NoStop}%
\bibitem [{\citenamefont {Biswas}\ \emph {et~al.}(2012)\citenamefont {Biswas},
  \citenamefont {He}, \citenamefont {Blum}, \citenamefont {Wu}, \citenamefont
  {Hogan}, \citenamefont {Seidman}, \citenamefont {Dravid},\ and\ \citenamefont
  {Kanatzidis}}]{biswasHighperformanceBulkThermoelectrics2012}%
  \BibitemOpen
  \bibfield  {author} {\bibinfo {author} {\bibfnamefont {K.}~\bibnamefont
  {Biswas}}, \bibinfo {author} {\bibfnamefont {J.}~\bibnamefont {He}}, \bibinfo
  {author} {\bibfnamefont {I.~D.}\ \bibnamefont {Blum}}, \bibinfo {author}
  {\bibfnamefont {C.-I.}\ \bibnamefont {Wu}}, \bibinfo {author} {\bibfnamefont
  {T.~P.}\ \bibnamefont {Hogan}}, \bibinfo {author} {\bibfnamefont {D.~N.}\
  \bibnamefont {Seidman}}, \bibinfo {author} {\bibfnamefont {V.~P.}\
  \bibnamefont {Dravid}}, \ and\ \bibinfo {author} {\bibfnamefont {M.~G.}\
  \bibnamefont {Kanatzidis}},\ }\href {\doibase 10.1038/nature11439} {\bibfield
   {journal} {\bibinfo  {journal} {Nature}\ }\textbf {\bibinfo {volume}
  {489}},\ \bibinfo {pages} {414} (\bibinfo {year} {2012})}\BibitemShut
  {NoStop}%
\bibitem [{\citenamefont {Behnia}\ and\ \citenamefont
  {Aubin}(2016)}]{behniaNernstEffectMetals2016}%
  \BibitemOpen
  \bibfield  {author} {\bibinfo {author} {\bibfnamefont {K.}~\bibnamefont
  {Behnia}}\ and\ \bibinfo {author} {\bibfnamefont {H.}~\bibnamefont {Aubin}},\
  }\href {\doibase 10.1088/0034-4885/79/4/046502} {\bibfield  {journal}
  {\bibinfo  {journal} {Rep. Prog. Phys.}\ }\textbf {\bibinfo {volume} {79}},\
  \bibinfo {pages} {046502} (\bibinfo {year} {2016})}\BibitemShut {NoStop}%
\bibitem [{\citenamefont {Watzman}\ \emph {et~al.}(2018)\citenamefont
  {Watzman}, \citenamefont {McCormick}, \citenamefont {Shekhar}, \citenamefont
  {Wu}, \citenamefont {Sun}, \citenamefont {Prakash}, \citenamefont {Felser},
  \citenamefont {Trivedi},\ and\ \citenamefont
  {Heremans}}]{watzmanDiracDispersionGenerates2018}%
  \BibitemOpen
  \bibfield  {author} {\bibinfo {author} {\bibfnamefont {S.~J.}\ \bibnamefont
  {Watzman}}, \bibinfo {author} {\bibfnamefont {T.~M.}\ \bibnamefont
  {McCormick}}, \bibinfo {author} {\bibfnamefont {C.}~\bibnamefont {Shekhar}},
  \bibinfo {author} {\bibfnamefont {S.-C.}\ \bibnamefont {Wu}}, \bibinfo
  {author} {\bibfnamefont {Y.}~\bibnamefont {Sun}}, \bibinfo {author}
  {\bibfnamefont {A.}~\bibnamefont {Prakash}}, \bibinfo {author} {\bibfnamefont
  {C.}~\bibnamefont {Felser}}, \bibinfo {author} {\bibfnamefont
  {N.}~\bibnamefont {Trivedi}}, \ and\ \bibinfo {author} {\bibfnamefont
  {J.~P.}\ \bibnamefont {Heremans}},\ }\href {\doibase
  10.1103/PhysRevB.97.161404} {\bibfield  {journal} {\bibinfo  {journal} {Phys.
  Rev. B}\ }\textbf {\bibinfo {volume} {97}},\ \bibinfo {pages} {161404}
  (\bibinfo {year} {2018})}\BibitemShut {NoStop}%
\bibitem [{\citenamefont {Fu}\ \emph {et~al.}(2018)\citenamefont {Fu},
  \citenamefont {Guin}, \citenamefont {Watzman}, \citenamefont {Li},
  \citenamefont {Liu}, \citenamefont {Kumar}, \citenamefont {S{\"u}{$\beta$}},
  \citenamefont {Schnelle}, \citenamefont {Auffermann}, \citenamefont
  {Shekhar}, \citenamefont {Sun}, \citenamefont {Gooth},\ and\ \citenamefont
  {Felser}}]{fuLargeNernstPower2018}%
  \BibitemOpen
  \bibfield  {author} {\bibinfo {author} {\bibfnamefont {C.}~\bibnamefont
  {Fu}}, \bibinfo {author} {\bibfnamefont {S.~N.}\ \bibnamefont {Guin}},
  \bibinfo {author} {\bibfnamefont {S.~J.}\ \bibnamefont {Watzman}}, \bibinfo
  {author} {\bibfnamefont {G.}~\bibnamefont {Li}}, \bibinfo {author}
  {\bibfnamefont {E.}~\bibnamefont {Liu}}, \bibinfo {author} {\bibfnamefont
  {N.}~\bibnamefont {Kumar}}, \bibinfo {author} {\bibfnamefont
  {V.}~\bibnamefont {S{\"u}{$\beta$}}}, \bibinfo {author} {\bibfnamefont
  {W.}~\bibnamefont {Schnelle}}, \bibinfo {author} {\bibfnamefont
  {G.}~\bibnamefont {Auffermann}}, \bibinfo {author} {\bibfnamefont
  {C.}~\bibnamefont {Shekhar}}, \bibinfo {author} {\bibfnamefont
  {Y.}~\bibnamefont {Sun}}, \bibinfo {author} {\bibfnamefont {J.}~\bibnamefont
  {Gooth}}, \ and\ \bibinfo {author} {\bibfnamefont {C.}~\bibnamefont
  {Felser}},\ }\href {\doibase 10.1039/C8EE02077A} {\bibfield  {journal}
  {\bibinfo  {journal} {Energy Environ. Sci.}\ }\textbf {\bibinfo {volume}
  {11}},\ \bibinfo {pages} {2813} (\bibinfo {year} {2018})}\BibitemShut
  {NoStop}%
\bibitem [{\citenamefont {Thierschmann}\ \emph {et~al.}(2015)\citenamefont
  {Thierschmann}, \citenamefont {S{\'a}nchez}, \citenamefont {Sothmann},
  \citenamefont {Arnold}, \citenamefont {Heyn}, \citenamefont {Hansen},
  \citenamefont {Buhmann},\ and\ \citenamefont
  {Molenkamp}}]{thierschmannThreeterminalEnergyHarvester2015}%
  \BibitemOpen
  \bibfield  {author} {\bibinfo {author} {\bibfnamefont {H.}~\bibnamefont
  {Thierschmann}}, \bibinfo {author} {\bibfnamefont {R.}~\bibnamefont
  {S{\'a}nchez}}, \bibinfo {author} {\bibfnamefont {B.}~\bibnamefont
  {Sothmann}}, \bibinfo {author} {\bibfnamefont {F.}~\bibnamefont {Arnold}},
  \bibinfo {author} {\bibfnamefont {C.}~\bibnamefont {Heyn}}, \bibinfo {author}
  {\bibfnamefont {W.}~\bibnamefont {Hansen}}, \bibinfo {author} {\bibfnamefont
  {H.}~\bibnamefont {Buhmann}}, \ and\ \bibinfo {author} {\bibfnamefont
  {L.~W.}\ \bibnamefont {Molenkamp}},\ }\href {\doibase 10.1038/nnano.2015.176}
  {\bibfield  {journal} {\bibinfo  {journal} {Nature Nanotech}\ }\textbf
  {\bibinfo {volume} {10}},\ \bibinfo {pages} {854} (\bibinfo {year}
  {2015})}\BibitemShut {NoStop}%
\bibitem [{\citenamefont {Skinner}\ and\ \citenamefont
  {Fu}(2018)}]{skinnerLargeNonsaturatingThermopower2018}%
  \BibitemOpen
  \bibfield  {author} {\bibinfo {author} {\bibfnamefont {B.}~\bibnamefont
  {Skinner}}\ and\ \bibinfo {author} {\bibfnamefont {L.}~\bibnamefont {Fu}},\
  }\href {\doibase 10.1126/sciadv.aat2621} {\bibfield  {journal} {\bibinfo
  {journal} {Science Advances}\ }\textbf {\bibinfo {volume} {4}},\ \bibinfo
  {pages} {eaat2621} (\bibinfo {year} {2018})}\BibitemShut {NoStop}%
\bibitem [{\citenamefont {Kozii}\ \emph {et~al.}(2019)\citenamefont {Kozii},
  \citenamefont {Skinner},\ and\ \citenamefont
  {Fu}}]{koziiThermoelectricHallConductivity2019}%
  \BibitemOpen
  \bibfield  {author} {\bibinfo {author} {\bibfnamefont {V.}~\bibnamefont
  {Kozii}}, \bibinfo {author} {\bibfnamefont {B.}~\bibnamefont {Skinner}}, \
  and\ \bibinfo {author} {\bibfnamefont {L.}~\bibnamefont {Fu}},\ }\href
  {\doibase 10.1103/PhysRevB.99.155123} {\bibfield  {journal} {\bibinfo
  {journal} {Phys. Rev. B}\ }\textbf {\bibinfo {volume} {99}},\ \bibinfo
  {pages} {155123} (\bibinfo {year} {2019})}\BibitemShut {NoStop}%
\bibitem [{\citenamefont {Fu}(2019)}]{fuCryogenicCoolingPower2019}%
  \BibitemOpen
  \bibfield  {author} {\bibinfo {author} {\bibfnamefont {L.}~\bibnamefont
  {Fu}},\ }\href@noop {} {\bibfield  {journal} {\bibinfo  {journal}
  {arXiv:1909.09506 [cond-mat]}\ } (\bibinfo {year} {2019})},\ \Eprint
  {http://arxiv.org/abs/1909.09506} {arXiv:1909.09506 [cond-mat]} \BibitemShut
  {NoStop}%
\bibitem [{\citenamefont {Boona}\ \emph {et~al.}(2014)\citenamefont {Boona},
  \citenamefont {Myers},\ and\ \citenamefont
  {Heremans}}]{boonaSpinCaloritronics2014}%
  \BibitemOpen
  \bibfield  {author} {\bibinfo {author} {\bibfnamefont {S.~R.}\ \bibnamefont
  {Boona}}, \bibinfo {author} {\bibfnamefont {R.~C.}\ \bibnamefont {Myers}}, \
  and\ \bibinfo {author} {\bibfnamefont {J.~P.}\ \bibnamefont {Heremans}},\
  }\href {\doibase 10.1039/C3EE43299H} {\bibfield  {journal} {\bibinfo
  {journal} {Energy Environ. Sci.}\ }\textbf {\bibinfo {volume} {7}},\ \bibinfo
  {pages} {885} (\bibinfo {year} {2014})}\BibitemShut {NoStop}%
\bibitem [{\citenamefont {Miyasato}\ \emph {et~al.}(2007)\citenamefont
  {Miyasato}, \citenamefont {Abe}, \citenamefont {Fujii}, \citenamefont
  {Asamitsu}, \citenamefont {Onoda}, \citenamefont {Onose}, \citenamefont
  {Nagaosa},\ and\ \citenamefont
  {Tokura}}]{miyasatoCrossoverBehaviorAnomalous2007}%
  \BibitemOpen
  \bibfield  {author} {\bibinfo {author} {\bibfnamefont {T.}~\bibnamefont
  {Miyasato}}, \bibinfo {author} {\bibfnamefont {N.}~\bibnamefont {Abe}},
  \bibinfo {author} {\bibfnamefont {T.}~\bibnamefont {Fujii}}, \bibinfo
  {author} {\bibfnamefont {A.}~\bibnamefont {Asamitsu}}, \bibinfo {author}
  {\bibfnamefont {S.}~\bibnamefont {Onoda}}, \bibinfo {author} {\bibfnamefont
  {Y.}~\bibnamefont {Onose}}, \bibinfo {author} {\bibfnamefont
  {N.}~\bibnamefont {Nagaosa}}, \ and\ \bibinfo {author} {\bibfnamefont
  {Y.}~\bibnamefont {Tokura}},\ }\href {\doibase 10.1103/PhysRevLett.99.086602}
  {\bibfield  {journal} {\bibinfo  {journal} {Phys. Rev. Lett.}\ }\textbf
  {\bibinfo {volume} {99}},\ \bibinfo {pages} {086602} (\bibinfo {year}
  {2007})}\BibitemShut {NoStop}%
\bibitem [{\citenamefont {Ikhlas}\ \emph {et~al.}(2017)\citenamefont {Ikhlas},
  \citenamefont {Tomita}, \citenamefont {Koretsune}, \citenamefont {Suzuki},
  \citenamefont {{Nishio-Hamane}}, \citenamefont {Arita}, \citenamefont
  {Otani},\ and\ \citenamefont {Nakatsuji}}]{ikhlasLargeAnomalousNernst2017}%
  \BibitemOpen
  \bibfield  {author} {\bibinfo {author} {\bibfnamefont {M.}~\bibnamefont
  {Ikhlas}}, \bibinfo {author} {\bibfnamefont {T.}~\bibnamefont {Tomita}},
  \bibinfo {author} {\bibfnamefont {T.}~\bibnamefont {Koretsune}}, \bibinfo
  {author} {\bibfnamefont {M.-T.}\ \bibnamefont {Suzuki}}, \bibinfo {author}
  {\bibfnamefont {D.}~\bibnamefont {{Nishio-Hamane}}}, \bibinfo {author}
  {\bibfnamefont {R.}~\bibnamefont {Arita}}, \bibinfo {author} {\bibfnamefont
  {Y.}~\bibnamefont {Otani}}, \ and\ \bibinfo {author} {\bibfnamefont
  {S.}~\bibnamefont {Nakatsuji}},\ }\href {\doibase 10.1038/nphys4181}
  {\bibfield  {journal} {\bibinfo  {journal} {Nature Physics}\ }\textbf
  {\bibinfo {volume} {13}},\ \bibinfo {pages} {1085} (\bibinfo {year}
  {2017})}\BibitemShut {NoStop}%
\bibitem [{\citenamefont {Liu}\ \emph {et~al.}(2018)\citenamefont {Liu},
  \citenamefont {Sun}, \citenamefont {Kumar}, \citenamefont {Muechler},
  \citenamefont {Sun}, \citenamefont {Jiao}, \citenamefont {Yang},
  \citenamefont {Liu}, \citenamefont {Liang}, \citenamefont {Xu}, \citenamefont
  {Kroder}, \citenamefont {S{\"u}{\ss}}, \citenamefont {Borrmann},
  \citenamefont {Shekhar}, \citenamefont {Wang}, \citenamefont {Xi},
  \citenamefont {Wang}, \citenamefont {Schnelle}, \citenamefont {Wirth},
  \citenamefont {Chen}, \citenamefont {Goennenwein},\ and\ \citenamefont
  {Felser}}]{liuGiantAnomalousHall2018}%
  \BibitemOpen
  \bibfield  {author} {\bibinfo {author} {\bibfnamefont {E.}~\bibnamefont
  {Liu}}, \bibinfo {author} {\bibfnamefont {Y.}~\bibnamefont {Sun}}, \bibinfo
  {author} {\bibfnamefont {N.}~\bibnamefont {Kumar}}, \bibinfo {author}
  {\bibfnamefont {L.}~\bibnamefont {Muechler}}, \bibinfo {author}
  {\bibfnamefont {A.}~\bibnamefont {Sun}}, \bibinfo {author} {\bibfnamefont
  {L.}~\bibnamefont {Jiao}}, \bibinfo {author} {\bibfnamefont {S.-Y.}\
  \bibnamefont {Yang}}, \bibinfo {author} {\bibfnamefont {D.}~\bibnamefont
  {Liu}}, \bibinfo {author} {\bibfnamefont {A.}~\bibnamefont {Liang}}, \bibinfo
  {author} {\bibfnamefont {Q.}~\bibnamefont {Xu}}, \bibinfo {author}
  {\bibfnamefont {J.}~\bibnamefont {Kroder}}, \bibinfo {author} {\bibfnamefont
  {V.}~\bibnamefont {S{\"u}{\ss}}}, \bibinfo {author} {\bibfnamefont
  {H.}~\bibnamefont {Borrmann}}, \bibinfo {author} {\bibfnamefont
  {C.}~\bibnamefont {Shekhar}}, \bibinfo {author} {\bibfnamefont
  {Z.}~\bibnamefont {Wang}}, \bibinfo {author} {\bibfnamefont {C.}~\bibnamefont
  {Xi}}, \bibinfo {author} {\bibfnamefont {W.}~\bibnamefont {Wang}}, \bibinfo
  {author} {\bibfnamefont {W.}~\bibnamefont {Schnelle}}, \bibinfo {author}
  {\bibfnamefont {S.}~\bibnamefont {Wirth}}, \bibinfo {author} {\bibfnamefont
  {Y.}~\bibnamefont {Chen}}, \bibinfo {author} {\bibfnamefont {S.~T.~B.}\
  \bibnamefont {Goennenwein}}, \ and\ \bibinfo {author} {\bibfnamefont
  {C.}~\bibnamefont {Felser}},\ }\href {\doibase 10.1038/s41567-018-0234-5}
  {\bibfield  {journal} {\bibinfo  {journal} {Nature Phys}\ }\textbf {\bibinfo
  {volume} {14}},\ \bibinfo {pages} {1125} (\bibinfo {year}
  {2018})}\BibitemShut {NoStop}%
\bibitem [{\citenamefont {Sakai}\ \emph {et~al.}(2018)\citenamefont {Sakai},
  \citenamefont {Mizuta}, \citenamefont {Nugroho}, \citenamefont {Sihombing},
  \citenamefont {Koretsune}, \citenamefont {Suzuki}, \citenamefont {Takemori},
  \citenamefont {Ishii}, \citenamefont {{Nishio-Hamane}}, \citenamefont
  {Arita}, \citenamefont {Goswami},\ and\ \citenamefont
  {Nakatsuji}}]{sakaiGiantAnomalousNernst2018}%
  \BibitemOpen
  \bibfield  {author} {\bibinfo {author} {\bibfnamefont {A.}~\bibnamefont
  {Sakai}}, \bibinfo {author} {\bibfnamefont {Y.~P.}\ \bibnamefont {Mizuta}},
  \bibinfo {author} {\bibfnamefont {A.~A.}\ \bibnamefont {Nugroho}}, \bibinfo
  {author} {\bibfnamefont {R.}~\bibnamefont {Sihombing}}, \bibinfo {author}
  {\bibfnamefont {T.}~\bibnamefont {Koretsune}}, \bibinfo {author}
  {\bibfnamefont {M.-T.}\ \bibnamefont {Suzuki}}, \bibinfo {author}
  {\bibfnamefont {N.}~\bibnamefont {Takemori}}, \bibinfo {author}
  {\bibfnamefont {R.}~\bibnamefont {Ishii}}, \bibinfo {author} {\bibfnamefont
  {D.}~\bibnamefont {{Nishio-Hamane}}}, \bibinfo {author} {\bibfnamefont
  {R.}~\bibnamefont {Arita}}, \bibinfo {author} {\bibfnamefont
  {P.}~\bibnamefont {Goswami}}, \ and\ \bibinfo {author} {\bibfnamefont
  {S.}~\bibnamefont {Nakatsuji}},\ }\href {\doibase 10.1038/s41567-018-0225-6}
  {\bibfield  {journal} {\bibinfo  {journal} {Nature Physics}\ }\textbf
  {\bibinfo {volume} {14}},\ \bibinfo {pages} {1119} (\bibinfo {year}
  {2018})}\BibitemShut {NoStop}%
\bibitem [{\citenamefont {Guin}\ \emph {et~al.}(2019)\citenamefont {Guin},
  \citenamefont {Vir}, \citenamefont {Zhang}, \citenamefont {Kumar},
  \citenamefont {Watzman}, \citenamefont {Fu}, \citenamefont {Liu},
  \citenamefont {Manna}, \citenamefont {Schnelle}, \citenamefont {Gooth},
  \citenamefont {Shekhar}, \citenamefont {Sun},\ and\ \citenamefont
  {Felser}}]{guinZeroFieldNernstEffect2019}%
  \BibitemOpen
  \bibfield  {author} {\bibinfo {author} {\bibfnamefont {S.~N.}\ \bibnamefont
  {Guin}}, \bibinfo {author} {\bibfnamefont {P.}~\bibnamefont {Vir}}, \bibinfo
  {author} {\bibfnamefont {Y.}~\bibnamefont {Zhang}}, \bibinfo {author}
  {\bibfnamefont {N.}~\bibnamefont {Kumar}}, \bibinfo {author} {\bibfnamefont
  {S.~J.}\ \bibnamefont {Watzman}}, \bibinfo {author} {\bibfnamefont
  {C.}~\bibnamefont {Fu}}, \bibinfo {author} {\bibfnamefont {E.}~\bibnamefont
  {Liu}}, \bibinfo {author} {\bibfnamefont {K.}~\bibnamefont {Manna}}, \bibinfo
  {author} {\bibfnamefont {W.}~\bibnamefont {Schnelle}}, \bibinfo {author}
  {\bibfnamefont {J.}~\bibnamefont {Gooth}}, \bibinfo {author} {\bibfnamefont
  {C.}~\bibnamefont {Shekhar}}, \bibinfo {author} {\bibfnamefont
  {Y.}~\bibnamefont {Sun}}, \ and\ \bibinfo {author} {\bibfnamefont
  {C.}~\bibnamefont {Felser}},\ }\href {\doibase 10.1002/adma.201806622}
  {\bibfield  {journal} {\bibinfo  {journal} {Advanced Materials}\ }\textbf
  {\bibinfo {volume} {31}},\ \bibinfo {pages} {1806622} (\bibinfo {year}
  {2019})}\BibitemShut {NoStop}%
\bibitem [{\citenamefont {Ding}\ \emph {et~al.}(2019)\citenamefont {Ding},
  \citenamefont {Koo}, \citenamefont {Xu}, \citenamefont {Li}, \citenamefont
  {Lu}, \citenamefont {Zhao}, \citenamefont {Wang}, \citenamefont {Yin},
  \citenamefont {Lei}, \citenamefont {Yan}, \citenamefont {Zhu},\ and\
  \citenamefont {Behnia}}]{dingIntrinsicAnomalousNernst2019}%
  \BibitemOpen
  \bibfield  {author} {\bibinfo {author} {\bibfnamefont {L.}~\bibnamefont
  {Ding}}, \bibinfo {author} {\bibfnamefont {J.}~\bibnamefont {Koo}}, \bibinfo
  {author} {\bibfnamefont {L.}~\bibnamefont {Xu}}, \bibinfo {author}
  {\bibfnamefont {X.}~\bibnamefont {Li}}, \bibinfo {author} {\bibfnamefont
  {X.}~\bibnamefont {Lu}}, \bibinfo {author} {\bibfnamefont {L.}~\bibnamefont
  {Zhao}}, \bibinfo {author} {\bibfnamefont {Q.}~\bibnamefont {Wang}}, \bibinfo
  {author} {\bibfnamefont {Q.}~\bibnamefont {Yin}}, \bibinfo {author}
  {\bibfnamefont {H.}~\bibnamefont {Lei}}, \bibinfo {author} {\bibfnamefont
  {B.}~\bibnamefont {Yan}}, \bibinfo {author} {\bibfnamefont {Z.}~\bibnamefont
  {Zhu}}, \ and\ \bibinfo {author} {\bibfnamefont {K.}~\bibnamefont {Behnia}},\
  }\href {\doibase 10.1103/PhysRevX.9.041061} {\bibfield  {journal} {\bibinfo
  {journal} {Phys. Rev. X}\ }\textbf {\bibinfo {volume} {9}},\ \bibinfo {pages}
  {041061} (\bibinfo {year} {2019})}\BibitemShut {NoStop}%
\bibitem [{\citenamefont {Xu}\ \emph {et~al.}(2019)\citenamefont {Xu},
  \citenamefont {Phelan},\ and\ \citenamefont
  {Chien}}]{xuLargeAnomalousNernst2019}%
  \BibitemOpen
  \bibfield  {author} {\bibinfo {author} {\bibfnamefont {J.}~\bibnamefont
  {Xu}}, \bibinfo {author} {\bibfnamefont {W.~A.}\ \bibnamefont {Phelan}}, \
  and\ \bibinfo {author} {\bibfnamefont {C.-L.}\ \bibnamefont {Chien}},\ }\href
  {\doibase 10.1021/acs.nanolett.9b03739} {\bibfield  {journal} {\bibinfo
  {journal} {Nano Lett.}\ }\textbf {\bibinfo {volume} {19}},\ \bibinfo {pages}
  {8250} (\bibinfo {year} {2019})}\BibitemShut {NoStop}%
\bibitem [{\citenamefont {Sakai}\ \emph {et~al.}(2020)\citenamefont {Sakai},
  \citenamefont {Minami}, \citenamefont {Koretsune}, \citenamefont {Chen},
  \citenamefont {Higo}, \citenamefont {Wang}, \citenamefont {Nomoto},
  \citenamefont {Hirayama}, \citenamefont {Miwa}, \citenamefont
  {{Nishio-Hamane}}, \citenamefont {Ishii}, \citenamefont {Arita},\ and\
  \citenamefont {Nakatsuji}}]{sakaiIronbasedBinaryFerromagnets2020}%
  \BibitemOpen
  \bibfield  {author} {\bibinfo {author} {\bibfnamefont {A.}~\bibnamefont
  {Sakai}}, \bibinfo {author} {\bibfnamefont {S.}~\bibnamefont {Minami}},
  \bibinfo {author} {\bibfnamefont {T.}~\bibnamefont {Koretsune}}, \bibinfo
  {author} {\bibfnamefont {T.}~\bibnamefont {Chen}}, \bibinfo {author}
  {\bibfnamefont {T.}~\bibnamefont {Higo}}, \bibinfo {author} {\bibfnamefont
  {Y.}~\bibnamefont {Wang}}, \bibinfo {author} {\bibfnamefont {T.}~\bibnamefont
  {Nomoto}}, \bibinfo {author} {\bibfnamefont {M.}~\bibnamefont {Hirayama}},
  \bibinfo {author} {\bibfnamefont {S.}~\bibnamefont {Miwa}}, \bibinfo {author}
  {\bibfnamefont {D.}~\bibnamefont {{Nishio-Hamane}}}, \bibinfo {author}
  {\bibfnamefont {F.}~\bibnamefont {Ishii}}, \bibinfo {author} {\bibfnamefont
  {R.}~\bibnamefont {Arita}}, \ and\ \bibinfo {author} {\bibfnamefont
  {S.}~\bibnamefont {Nakatsuji}},\ }\href {\doibase 10.1038/s41586-020-2230-z}
  {\bibfield  {journal} {\bibinfo  {journal} {Nature}\ }\textbf {\bibinfo
  {volume} {581}},\ \bibinfo {pages} {53} (\bibinfo {year} {2020})}\BibitemShut
  {NoStop}%
\bibitem [{\citenamefont {Xiao}\ \emph {et~al.}(2010)\citenamefont {Xiao},
  \citenamefont {Chang},\ and\ \citenamefont
  {Niu}}]{xiaoBerryPhaseEffects2010}%
  \BibitemOpen
  \bibfield  {author} {\bibinfo {author} {\bibfnamefont {D.}~\bibnamefont
  {Xiao}}, \bibinfo {author} {\bibfnamefont {M.-C.}\ \bibnamefont {Chang}}, \
  and\ \bibinfo {author} {\bibfnamefont {Q.}~\bibnamefont {Niu}},\ }\href
  {\doibase 10.1103/RevModPhys.82.1959} {\bibfield  {journal} {\bibinfo
  {journal} {Rev. Mod. Phys.}\ }\textbf {\bibinfo {volume} {82}},\ \bibinfo
  {pages} {1959} (\bibinfo {year} {2010})}\BibitemShut {NoStop}%
\bibitem [{\citenamefont {Nagaosa}\ \emph {et~al.}(2010)\citenamefont
  {Nagaosa}, \citenamefont {Sinova}, \citenamefont {Onoda}, \citenamefont
  {MacDonald},\ and\ \citenamefont {Ong}}]{nagaosaAnomalousHallEffect2010}%
  \BibitemOpen
  \bibfield  {author} {\bibinfo {author} {\bibfnamefont {N.}~\bibnamefont
  {Nagaosa}}, \bibinfo {author} {\bibfnamefont {J.}~\bibnamefont {Sinova}},
  \bibinfo {author} {\bibfnamefont {S.}~\bibnamefont {Onoda}}, \bibinfo
  {author} {\bibfnamefont {A.~H.}\ \bibnamefont {MacDonald}}, \ and\ \bibinfo
  {author} {\bibfnamefont {N.~P.}\ \bibnamefont {Ong}},\ }\href {\doibase
  10.1103/RevModPhys.82.1539} {\bibfield  {journal} {\bibinfo  {journal} {Rev.
  Mod. Phys.}\ }\textbf {\bibinfo {volume} {82}},\ \bibinfo {pages} {1539}
  (\bibinfo {year} {2010})}\BibitemShut {NoStop}%
\bibitem [{\citenamefont {Lundgren}\ \emph {et~al.}(2014)\citenamefont
  {Lundgren}, \citenamefont {Laurell},\ and\ \citenamefont
  {Fiete}}]{lundgrenThermoelectricPropertiesWeyl2014}%
  \BibitemOpen
  \bibfield  {author} {\bibinfo {author} {\bibfnamefont {R.}~\bibnamefont
  {Lundgren}}, \bibinfo {author} {\bibfnamefont {P.}~\bibnamefont {Laurell}}, \
  and\ \bibinfo {author} {\bibfnamefont {G.~A.}\ \bibnamefont {Fiete}},\ }\href
  {\doibase 10.1103/PhysRevB.90.165115} {\bibfield  {journal} {\bibinfo
  {journal} {Phys. Rev. B}\ }\textbf {\bibinfo {volume} {90}},\ \bibinfo
  {pages} {165115} (\bibinfo {year} {2014})}\BibitemShut {NoStop}%
\bibitem [{\citenamefont {Ye}\ \emph {et~al.}(2018)\citenamefont {Ye},
  \citenamefont {Kang}, \citenamefont {Liu}, \citenamefont {von Cube},
  \citenamefont {Wicker}, \citenamefont {Suzuki}, \citenamefont {Jozwiak},
  \citenamefont {Bostwick}, \citenamefont {Rotenberg}, \citenamefont {Bell},
  \citenamefont {Fu}, \citenamefont {Comin},\ and\ \citenamefont
  {Checkelsky}}]{yeMassiveDiracFermions2018}%
  \BibitemOpen
  \bibfield  {author} {\bibinfo {author} {\bibfnamefont {L.}~\bibnamefont
  {Ye}}, \bibinfo {author} {\bibfnamefont {M.}~\bibnamefont {Kang}}, \bibinfo
  {author} {\bibfnamefont {J.}~\bibnamefont {Liu}}, \bibinfo {author}
  {\bibfnamefont {F.}~\bibnamefont {von Cube}}, \bibinfo {author}
  {\bibfnamefont {C.~R.}\ \bibnamefont {Wicker}}, \bibinfo {author}
  {\bibfnamefont {T.}~\bibnamefont {Suzuki}}, \bibinfo {author} {\bibfnamefont
  {C.}~\bibnamefont {Jozwiak}}, \bibinfo {author} {\bibfnamefont
  {A.}~\bibnamefont {Bostwick}}, \bibinfo {author} {\bibfnamefont
  {E.}~\bibnamefont {Rotenberg}}, \bibinfo {author} {\bibfnamefont {D.~C.}\
  \bibnamefont {Bell}}, \bibinfo {author} {\bibfnamefont {L.}~\bibnamefont
  {Fu}}, \bibinfo {author} {\bibfnamefont {R.}~\bibnamefont {Comin}}, \ and\
  \bibinfo {author} {\bibfnamefont {J.~G.}\ \bibnamefont {Checkelsky}},\ }\href
  {\doibase 10.1038/nature25987} {\bibfield  {journal} {\bibinfo  {journal}
  {Nature}\ }\textbf {\bibinfo {volume} {555}},\ \bibinfo {pages} {638}
  (\bibinfo {year} {2018})}\BibitemShut {NoStop}%
\bibitem [{\citenamefont {Sodemann}\ and\ \citenamefont
  {Fu}(2015)}]{sodemannQuantumNonlinearHall2015}%
  \BibitemOpen
  \bibfield  {author} {\bibinfo {author} {\bibfnamefont {I.}~\bibnamefont
  {Sodemann}}\ and\ \bibinfo {author} {\bibfnamefont {L.}~\bibnamefont {Fu}},\
  }\href {\doibase 10.1103/PhysRevLett.115.216806} {\bibfield  {journal}
  {\bibinfo  {journal} {Phys. Rev. Lett.}\ }\textbf {\bibinfo {volume} {115}},\
  \bibinfo {pages} {216806} (\bibinfo {year} {2015})}\BibitemShut {NoStop}%
\bibitem [{\citenamefont {Ma}\ \emph {et~al.}(2019)\citenamefont {Ma},
  \citenamefont {Xu}, \citenamefont {Shen}, \citenamefont {MacNeill},
  \citenamefont {Fatemi}, \citenamefont {Chang}, \citenamefont {Valdivia},
  \citenamefont {Wu}, \citenamefont {Du}, \citenamefont {Hsu}, \citenamefont
  {Fang}, \citenamefont {Gibson}, \citenamefont {Watanabe}, \citenamefont
  {Taniguchi}, \citenamefont {Cava}, \citenamefont {Kaxiras}, \citenamefont
  {Lu}, \citenamefont {Lin}, \citenamefont {Fu}, \citenamefont {Gedik},\ and\
  \citenamefont {{Jarillo-Herrero}}}]{maObservationNonlinearHall2019}%
  \BibitemOpen
  \bibfield  {author} {\bibinfo {author} {\bibfnamefont {Q.}~\bibnamefont
  {Ma}}, \bibinfo {author} {\bibfnamefont {S.-Y.}\ \bibnamefont {Xu}}, \bibinfo
  {author} {\bibfnamefont {H.}~\bibnamefont {Shen}}, \bibinfo {author}
  {\bibfnamefont {D.}~\bibnamefont {MacNeill}}, \bibinfo {author}
  {\bibfnamefont {V.}~\bibnamefont {Fatemi}}, \bibinfo {author} {\bibfnamefont
  {T.-R.}\ \bibnamefont {Chang}}, \bibinfo {author} {\bibfnamefont {A.~M.~M.}\
  \bibnamefont {Valdivia}}, \bibinfo {author} {\bibfnamefont {S.}~\bibnamefont
  {Wu}}, \bibinfo {author} {\bibfnamefont {Z.}~\bibnamefont {Du}}, \bibinfo
  {author} {\bibfnamefont {C.-H.}\ \bibnamefont {Hsu}}, \bibinfo {author}
  {\bibfnamefont {S.}~\bibnamefont {Fang}}, \bibinfo {author} {\bibfnamefont
  {Q.~D.}\ \bibnamefont {Gibson}}, \bibinfo {author} {\bibfnamefont
  {K.}~\bibnamefont {Watanabe}}, \bibinfo {author} {\bibfnamefont
  {T.}~\bibnamefont {Taniguchi}}, \bibinfo {author} {\bibfnamefont {R.~J.}\
  \bibnamefont {Cava}}, \bibinfo {author} {\bibfnamefont {E.}~\bibnamefont
  {Kaxiras}}, \bibinfo {author} {\bibfnamefont {H.-Z.}\ \bibnamefont {Lu}},
  \bibinfo {author} {\bibfnamefont {H.}~\bibnamefont {Lin}}, \bibinfo {author}
  {\bibfnamefont {L.}~\bibnamefont {Fu}}, \bibinfo {author} {\bibfnamefont
  {N.}~\bibnamefont {Gedik}}, \ and\ \bibinfo {author} {\bibfnamefont
  {P.}~\bibnamefont {{Jarillo-Herrero}}},\ }\href {\doibase
  10.1038/s41586-018-0807-6} {\bibfield  {journal} {\bibinfo  {journal}
  {Nature}\ }\textbf {\bibinfo {volume} {565}},\ \bibinfo {pages} {337}
  (\bibinfo {year} {2019})}\BibitemShut {NoStop}%
\bibitem [{\citenamefont {Du}\ \emph {et~al.}(2019)\citenamefont {Du},
  \citenamefont {Wang}, \citenamefont {Li}, \citenamefont {Lu},\ and\
  \citenamefont {Xie}}]{duDisorderinducedNonlinearHall2019}%
  \BibitemOpen
  \bibfield  {author} {\bibinfo {author} {\bibfnamefont {Z.~Z.}\ \bibnamefont
  {Du}}, \bibinfo {author} {\bibfnamefont {C.~M.}\ \bibnamefont {Wang}},
  \bibinfo {author} {\bibfnamefont {S.}~\bibnamefont {Li}}, \bibinfo {author}
  {\bibfnamefont {H.-Z.}\ \bibnamefont {Lu}}, \ and\ \bibinfo {author}
  {\bibfnamefont {X.~C.}\ \bibnamefont {Xie}},\ }\href {\doibase
  10.1038/s41467-019-10941-3} {\bibfield  {journal} {\bibinfo  {journal} {Nat
  Commun}\ }\textbf {\bibinfo {volume} {10}},\ \bibinfo {pages} {1} (\bibinfo
  {year} {2019})}\BibitemShut {NoStop}%
\bibitem [{\citenamefont {Isobe}\ \emph {et~al.}(2020)\citenamefont {Isobe},
  \citenamefont {Xu},\ and\ \citenamefont
  {Fu}}]{isobeHighfrequencyRectificationChiral2020}%
  \BibitemOpen
  \bibfield  {author} {\bibinfo {author} {\bibfnamefont {H.}~\bibnamefont
  {Isobe}}, \bibinfo {author} {\bibfnamefont {S.-Y.}\ \bibnamefont {Xu}}, \
  and\ \bibinfo {author} {\bibfnamefont {L.}~\bibnamefont {Fu}},\ }\href
  {\doibase 10.1126/sciadv.aay2497} {\bibfield  {journal} {\bibinfo  {journal}
  {Science Advances}\ }\textbf {\bibinfo {volume} {6}},\ \bibinfo {pages}
  {eaay2497} (\bibinfo {year} {2020})}\BibitemShut {NoStop}%
\bibitem [{\citenamefont {Papaj}\ and\ \citenamefont
  {Fu}(2019)}]{papajMagnusHallEffect2019}%
  \BibitemOpen
  \bibfield  {author} {\bibinfo {author} {\bibfnamefont {M.}~\bibnamefont
  {Papaj}}\ and\ \bibinfo {author} {\bibfnamefont {L.}~\bibnamefont {Fu}},\
  }\href {\doibase 10.1103/PhysRevLett.123.216802} {\bibfield  {journal}
  {\bibinfo  {journal} {Phys. Rev. Lett.}\ }\textbf {\bibinfo {volume} {123}},\
  \bibinfo {pages} {216802} (\bibinfo {year} {2019})}\BibitemShut {NoStop}%
\bibitem [{\citenamefont {Sinitsyn}\ \emph {et~al.}(2006)\citenamefont
  {Sinitsyn}, \citenamefont {Niu},\ and\ \citenamefont
  {MacDonald}}]{sinitsynCoordinateShiftSemiclassical2006}%
  \BibitemOpen
  \bibfield  {author} {\bibinfo {author} {\bibfnamefont {N.~A.}\ \bibnamefont
  {Sinitsyn}}, \bibinfo {author} {\bibfnamefont {Q.}~\bibnamefont {Niu}}, \
  and\ \bibinfo {author} {\bibfnamefont {A.~H.}\ \bibnamefont {MacDonald}},\
  }\href {\doibase 10.1103/PhysRevB.73.075318} {\bibfield  {journal} {\bibinfo
  {journal} {Phys. Rev. B}\ }\textbf {\bibinfo {volume} {73}},\ \bibinfo
  {pages} {075318} (\bibinfo {year} {2006})}\BibitemShut {NoStop}%
\bibitem [{\citenamefont {Xiao}\ \emph {et~al.}(2006)\citenamefont {Xiao},
  \citenamefont {Yao}, \citenamefont {Fang},\ and\ \citenamefont
  {Niu}}]{xiaoBerryPhaseEffectAnomalous2006}%
  \BibitemOpen
  \bibfield  {author} {\bibinfo {author} {\bibfnamefont {D.}~\bibnamefont
  {Xiao}}, \bibinfo {author} {\bibfnamefont {Y.}~\bibnamefont {Yao}}, \bibinfo
  {author} {\bibfnamefont {Z.}~\bibnamefont {Fang}}, \ and\ \bibinfo {author}
  {\bibfnamefont {Q.}~\bibnamefont {Niu}},\ }\href {\doibase
  10.1103/PhysRevLett.97.026603} {\bibfield  {journal} {\bibinfo  {journal}
  {Phys. Rev. Lett.}\ }\textbf {\bibinfo {volume} {97}},\ \bibinfo {pages}
  {026603} (\bibinfo {year} {2006})}\BibitemShut {NoStop}%
\bibitem [{\citenamefont {Sinitsyn}\ \emph {et~al.}(2007)\citenamefont
  {Sinitsyn}, \citenamefont {MacDonald}, \citenamefont {Jungwirth},
  \citenamefont {Dugaev},\ and\ \citenamefont
  {Sinova}}]{sinitsynAnomalousHallEffect2007}%
  \BibitemOpen
  \bibfield  {author} {\bibinfo {author} {\bibfnamefont {N.~A.}\ \bibnamefont
  {Sinitsyn}}, \bibinfo {author} {\bibfnamefont {A.~H.}\ \bibnamefont
  {MacDonald}}, \bibinfo {author} {\bibfnamefont {T.}~\bibnamefont
  {Jungwirth}}, \bibinfo {author} {\bibfnamefont {V.~K.}\ \bibnamefont
  {Dugaev}}, \ and\ \bibinfo {author} {\bibfnamefont {J.}~\bibnamefont
  {Sinova}},\ }\href {\doibase 10.1103/PhysRevB.75.045315} {\bibfield
  {journal} {\bibinfo  {journal} {Phys. Rev. B}\ }\textbf {\bibinfo {volume}
  {75}},\ \bibinfo {pages} {045315} (\bibinfo {year} {2007})}\BibitemShut
  {NoStop}%
\bibitem [{\citenamefont {Ye}\ \emph {et~al.}(2019)\citenamefont {Ye},
  \citenamefont {Chan}, \citenamefont {McDonald}, \citenamefont {Graf},
  \citenamefont {Kang}, \citenamefont {Liu}, \citenamefont {Suzuki},
  \citenamefont {Comin}, \citenamefont {Fu},\ and\ \citenamefont
  {Checkelsky}}]{yeHaasvanAlphenEffect2019a}%
  \BibitemOpen
  \bibfield  {author} {\bibinfo {author} {\bibfnamefont {L.}~\bibnamefont
  {Ye}}, \bibinfo {author} {\bibfnamefont {M.~K.}\ \bibnamefont {Chan}},
  \bibinfo {author} {\bibfnamefont {R.~D.}\ \bibnamefont {McDonald}}, \bibinfo
  {author} {\bibfnamefont {D.}~\bibnamefont {Graf}}, \bibinfo {author}
  {\bibfnamefont {M.}~\bibnamefont {Kang}}, \bibinfo {author} {\bibfnamefont
  {J.}~\bibnamefont {Liu}}, \bibinfo {author} {\bibfnamefont {T.}~\bibnamefont
  {Suzuki}}, \bibinfo {author} {\bibfnamefont {R.}~\bibnamefont {Comin}},
  \bibinfo {author} {\bibfnamefont {L.}~\bibnamefont {Fu}}, \ and\ \bibinfo
  {author} {\bibfnamefont {J.~G.}\ \bibnamefont {Checkelsky}},\ }\href
  {\doibase 10.1038/s41467-019-12822-1} {\bibfield  {journal} {\bibinfo
  {journal} {Nat Commun}\ }\textbf {\bibinfo {volume} {10}},\ \bibinfo {pages}
  {1} (\bibinfo {year} {2019})}\BibitemShut {NoStop}%
\bibitem [{\citenamefont {Burkov}(2014)}]{burkovAnomalousHallEffect2014}%
  \BibitemOpen
  \bibfield  {author} {\bibinfo {author} {\bibfnamefont {A.~A.}\ \bibnamefont
  {Burkov}},\ }\href {\doibase 10.1103/PhysRevLett.113.187202} {\bibfield
  {journal} {\bibinfo  {journal} {Phys. Rev. Lett.}\ }\textbf {\bibinfo
  {volume} {113}},\ \bibinfo {pages} {187202} (\bibinfo {year}
  {2014})}\BibitemShut {NoStop}%
\bibitem [{\citenamefont {Ferreiros}\ \emph {et~al.}(2017)\citenamefont
  {Ferreiros}, \citenamefont {Zyuzin},\ and\ \citenamefont
  {Bardarson}}]{ferreirosAnomalousNernstThermal2017}%
  \BibitemOpen
  \bibfield  {author} {\bibinfo {author} {\bibfnamefont {Y.}~\bibnamefont
  {Ferreiros}}, \bibinfo {author} {\bibfnamefont {A.~A.}\ \bibnamefont
  {Zyuzin}}, \ and\ \bibinfo {author} {\bibfnamefont {J.~H.}\ \bibnamefont
  {Bardarson}},\ }\href {\doibase 10.1103/PhysRevB.96.115202} {\bibfield
  {journal} {\bibinfo  {journal} {Phys. Rev. B}\ }\textbf {\bibinfo {volume}
  {96}},\ \bibinfo {pages} {115202} (\bibinfo {year} {2017})}\BibitemShut
  {NoStop}%
\bibitem [{\citenamefont {Morali}\ \emph {et~al.}(2019)\citenamefont {Morali},
  \citenamefont {Batabyal}, \citenamefont {Nag}, \citenamefont {Liu},
  \citenamefont {Xu}, \citenamefont {Sun}, \citenamefont {Yan}, \citenamefont
  {Felser}, \citenamefont {Avraham},\ and\ \citenamefont
  {Beidenkopf}}]{moraliFermiarcDiversitySurface2019}%
  \BibitemOpen
  \bibfield  {author} {\bibinfo {author} {\bibfnamefont {N.}~\bibnamefont
  {Morali}}, \bibinfo {author} {\bibfnamefont {R.}~\bibnamefont {Batabyal}},
  \bibinfo {author} {\bibfnamefont {P.~K.}\ \bibnamefont {Nag}}, \bibinfo
  {author} {\bibfnamefont {E.}~\bibnamefont {Liu}}, \bibinfo {author}
  {\bibfnamefont {Q.}~\bibnamefont {Xu}}, \bibinfo {author} {\bibfnamefont
  {Y.}~\bibnamefont {Sun}}, \bibinfo {author} {\bibfnamefont {B.}~\bibnamefont
  {Yan}}, \bibinfo {author} {\bibfnamefont {C.}~\bibnamefont {Felser}},
  \bibinfo {author} {\bibfnamefont {N.}~\bibnamefont {Avraham}}, \ and\
  \bibinfo {author} {\bibfnamefont {H.}~\bibnamefont {Beidenkopf}},\ }\href
  {\doibase 10.1126/science.aav2334} {\bibfield  {journal} {\bibinfo  {journal}
  {Science}\ }\textbf {\bibinfo {volume} {365}},\ \bibinfo {pages} {1286}
  (\bibinfo {year} {2019})}\BibitemShut {NoStop}%
\end{thebibliography}%

\end{document}